\documentclass[11pt]{article}
\usepackage{graphicx}
\usepackage{amssymb}
\RequirePackage{mathptmx}
\RequirePackage[T1]{fontenc}
\usepackage{heppennames2}
\usepackage{lhchiggs}
\usepackage{cernunits}
\DeclareSymbolFont{forjmath}{OT1}{cmr}{m}{sl}
\DeclareMathSymbol{\Jmath}{\mathord}{forjmath}{'021}
\def\jmath{\Jmath}
\DeclareFontFamily{OT1}{cmr}{}
\DeclareFontFamily{OT1}{cmss}{}
\usepackage{pifont}
\usepackage{vicent}
\usepackage{parskip}
\usepackage{enumerate}

\allowdisplaybreaks
\def\csumb{Dipartimento di Fisica Teorica, Universit\`a di Torino, Italy\\
           INFN, Sezione di Torino, Italy}
\def\support{\footnote{Work supported by MIUR under contract
    2001023713$\_$006 and by UniTo - Compagnia di San Paolo under contract ORTO11TPXK.}}
\def\Title#1{\begin{center} {\Large\bf #1 } \end{center}}

\def\Author#1{\begin{center}{ \sc #1} \end{center}}
\def\Address#1{\begin{center}{ \it #1} \end{center}}

\newenvironment{Abstract}{\begin{quotation}  }{\end{quotation}}

\def\Acknowledgments{\bigskip  \bigskip \begin{center}
          \large\bf Acknowledgments\end{center}}
\def\email#1{\footnote{#1}}
\makeatletter
\def\section{\@startsection{section}{0}{\z@}{5.5ex plus .5ex minus
 1.5ex}{2.3ex plus .2ex}{\large\bf}}
\def\subsection{\@startsection{subsection}{1}{\z@}{3.5ex plus .5ex minus
 1.5ex}{1.3ex plus .2ex}{\normalsize\bf}}
\def\subsubsection{\@startsection{subsubsection}{2}{\z@}{-3.5ex plus
-1ex minus  -.2ex}{2.3ex plus .2ex}{\normalsize\sl}}

\renewcommand{\@makecaption}[2]{%
   \vskip 10pt
   \setbox\@tempboxa\hbox{\small #1: #2}
   \ifdim \wd\@tempboxa >\hsize     
       \small #1: #2\par          
     \else                        
       \hbox to\hsize{\hfil\box\@tempboxa\hfil}
   \fi}
 \def\citenum#1{{\def\@cite##1##2{##1}\cite{#1}}}
\def\citea#1{\@cite{#1}{}}
\newcount\@tempcntc
\def\@citex[#1]#2{\if@filesw\immediate\write\@auxout{\string\citation{#2}}\fi
  \@tempcnta\z@\@tempcntb\m@ne\def\@citea{}\@cite{\@for\@citeb:=#2\do
    {\@ifundefined
       {b@\@citeb}{\@citeo\@tempcntb\m@ne\@citea\def\@citea{,}{\bf }\@warning
       {Citation `\@citeb' on page \thepage \space undefined}}%
    {\setbox\z@\hbox{\global\@tempcntc0\csname b@\@citeb\endcsname\relax}%
     \ifnum\@tempcntc=\z@ \@citeo\@tempcntb\m@ne
       \@citea\def\@citea{,}\hbox{\csname b@\@citeb\endcsname}%
     \else
      \advance\@tempcntb\@ne
      \ifnum\@tempcntb=\@tempcntc
      \else\advance\@tempcntb\m@ne\@citeo
      \@tempcnta\@tempcntc\@tempcntb\@tempcntc\fi\fi}}\@citeo}{#1}}
\def\@citeo{\ifnum\@tempcnta>\@tempcntb\else\@citea\def\@citea{,}%
  \ifnum\@tempcnta=\@tempcntb\the\@tempcnta\else
  {\advance\@tempcnta\@ne\ifnum\@tempcnta=\@tempcntb \else\def\@citea{--}\fi
    \advance\@tempcnta\m@ne\the\@tempcnta\@citea\the\@tempcntb}\fi\fi}
\makeatother
\DeclareRobustCommand{\PA}{\HepParticle{A}{}{}\Xspace}
\DeclareRobustCommand{\PV}{\HepParticle{V}{}{}\Xspace}
\DeclareRobustCommand{\PX}{\HepParticle{X}{}{}\Xspace}
\DeclareRobustCommand{\Pf}{\HepParticle{f}{}{}\Xspace}

\DeclareRobustCommand{\PF}{\HepParticle{F}{}{}\Xspace}

\newcommand{\myLO}{\mathrm{\scriptscriptstyle{LO}}}

\newcommand{\myNNLO}{\mathrm{\scriptscriptstyle{NNLO}}}

\newcommand{\mySM}{\rm{\scriptscriptstyle{SM}}}

\newcommand{\ssA}{{\mathrm{A}}}
\newcommand{\ssB}{{\mathrm{B}}}

\newcommand{\ssF}{{\mathrm{F}}}
\newcommand{\ssR}{{\mathrm{R}}}
\newcommand{\ssD}{{\mathrm{D}}}
\newcommand{\ssI}{{\mathrm{I}}}

\newcommand{\ssS}{{\mathrm{S}}}

\newcommand{\ssM}{{\mathrm{M}}}

\newcommand{\sH}{\mathrm{H}}

\newcommand{\ssZZ}{{\scriptscriptstyle{\PZ\PZ}}}

\newcommand{\ssZ}{{\mathrm{Z}}}
\newcommand{\bqas}{\begin{eqnarray*}}
\newcommand{\eqas}{\end{eqnarray*}}
\newcommand{\nl}{\nonumber\\}

\newcommand{\lpar}{\left(}                            
\newcommand{\rpar}{\right)}

\newcommand{\bq}{\begin{equation}}                    
\newcommand{\eq}{\end{equation}}
\newcommand{\bqa}{\arraycolsep 0.14em\begin{eqnarray}}
\newcommand{\eqa}{\end{eqnarray}}
\newcommand{\ba}[1]{\begin{array}{#1}}
\newcommand{\ea}{\end{array}}
\newcommand{\ben}{\begin{enumerate}}
\newcommand{\een}{\end{enumerate}}
\newcommand{\bei}{\begin{itemize}}
\newcommand{\eei}{\end{itemize}}
\newcommand{\eqn}[1]{Eq.(\ref{#1})}

\newcommand{\eqnsc}[2]{Eqs.(\ref{#1}) and (\ref{#2})}

\newcommand{\hatp}{{\hat p}}

\newcommand{\bmid}{\Bigr|}

\newcommand{\Bref}[1]{Ref.~\cite{#1}}
\newcommand{\Brefs}[1]{Refs.~\cite{#1}}

\newcommand{\eg}{e.g.\xspace}
\newcommand{\ie}{i.e.\xspace}
\newcommand{\etc}{etc.\@\xspace}

\newcommand{\mh}{\mathswitch {M_{\PH}}}

\newcommand{\mBh}{\mathswitch {{\overline M}_{\PH}}}
\newcommand{\mBhs}{\mathswitch {{\overline M}^2_{\PH}}}

\newcommand{\cph}{\mathswitch {s_{\PH}}}

\newcommand{\muh}{\mathswitch {\mu_{\PH}}}

\newcommand{\muhs}{\mathswitch {\mu^2_{\PH}}}
\newcommand{\omuhs}{\mathswitch {{\hat\mu}^2_{\PH}}}
\newcommand{\gh}{\mathswitch {\gamma_{\PH}}}

\newcommand{\GOL}{\mathswitch {\overline\Gamma}}

\newcommand{\muR}{\mathswitch {\mu_{\ssR}}}
\newcommand{\muF}{\mathswitch {\mu_{\ssF}}}
\newcommand{\muRs}{\mathswitch {\mu^2_{\ssR}}}
\newcommand{\muFs}{\mathswitch {\mu^2_{\ssF}}}
\newcommand{\tot}{{\mbox{\scriptsize tot}}}
\newcommand{\myprod}{{\mbox{\scriptsize prod}}}
\newcommand{\prop}{{\mbox{\scriptsize prop}}}
\newcommand{\dec}{{\mbox{\scriptsize dec}}}
\newcommand{\off}{{\mbox{\scriptsize off}}}

\newcommand{\bck}{{\mbox{\scriptsize NR}}}
\newcommand{\res}{{\mbox{\scriptsize R}}}
\newcommand{\mysoft}{{\mbox{\scriptsize soft}}}

\newcommand{\peak}{{\mbox{\scriptsize peak}}}
\newcommand{\eff}{{\mbox{\scriptsize eff}}}
\newcommand{\rest}{{\mbox{\scriptsize rest}}}

\newcommand{\ac}{{\mbox{\scriptsize all}}}

\newcommand{\Ds}{{\mathrm{D}}}
\newcommand{\mrS}{{\mathrm{S}}}
\newcommand{\mrB}{{\mathrm{B}}}
\newcommand{\mrI}{{\mathrm{I}}}
\newcommand{\Kfac}[2]{\mathrm{K}^{#1}_{#2}}
\newcommand{\Kf}{\mathrm{K}}
\newcommand{\SpI}{\mathrm{S} + \mathrm{I}}
\newcommand{\mfs}{\mathswitch {M_{\Pf}}}
\newcommand{\cmark}{\ding{51}\Xspace}
\newcommand{\ccmark}{\ding{55}\Xspace}

\newtheorem{theorem}{Theorem}[section]

\newtheorem{proposition}[theorem]{Proposition}

\newenvironment{definition}[1][Definition]{\begin{trivlist}
\item[\hskip \labelsep {\bfseries #1}]}{\end{trivlist}}

\newenvironment{remark}[1][Remark]{\begin{trivlist}
\item[\hskip \labelsep {\bfseries #1}]}{\end{trivlist}}
\begin{document}
\begin{titlepage}
%
\vfill
\def\thefootnote{\fnsymbol{footnote}}
\Title{\LARGE \sffamily \bfseries
Higgs CAT\support}
\vspace{1.cm}
\Author{\normalsize \bfseries \sffamily
Giampiero Passarino \email{giampiero@to.infn.it}}               
\Address{\csumb}
\vspace{2.cm}
\begin{Abstract}
\noindent 
Higgs Computed Axial Tomography, an excerpt.
Taking a closer look at the camel-shaped tail of the light Higgs boson resonance and looking to
the transformation of the (camel-shaped) signal into a square-root--shaped signal + interference
\footnote{$\dots$ \begin{itshape}
and the devil hath power to assume a pleasing shape
\end{itshape}, Hamlet, Act II, scene 2} with particular emphasis on residual theoretical 
uncertainties.
\end{Abstract}
\includegraphics[width=3.0cm]{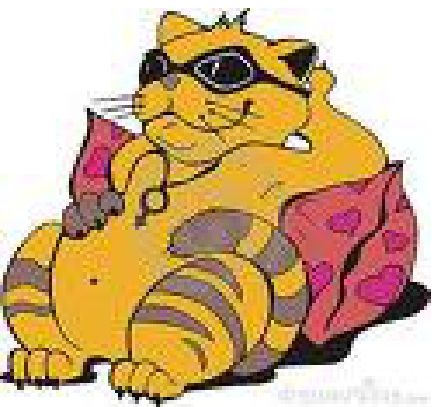}

\vfill
\end{titlepage}
\def\thefootnote{\arabic{footnote}}
\setcounter{footnote}{0}
\small
\thispagestyle{empty}
\tableofcontents
\normalsize
\clearpage
\setcounter{page}{1}
\section{Introduction \label{intro}}
Somebody had an idea, somebody else gave it wings, a third group did the cut-and-count, and 
a fourth did a shape-based analysis\footnote{Inspired by a friend}.
Ideas are like rabbits. You get a couple, learn how to handle them, and pretty soon you have a 
dozen. 

Here I present a few personal recollections and observations on what is necessary in order 
to obtain the most accurate theoretical predictions outside the Higgs-like resonance region,
given the present level of knowledge.
\section{An old idea\label{OI}}
The problem of determining resonance parameters in $\Pep \Pem$ annihilation, including
initial state radiative corrections and resolution corrections is an old one, see
\Bref{Jackson:1975vf}. For the interested reader we recommend the original
\Brefs{Jackson:1975vf,Alexander:1987be} or the summary in Chap.~2 of \Bref{Bardin:1999ak}.
\subsection{Higgs intrinsic width \label{Hiw}}
Is there anything we can say about what the intrinsic width of the light resonance is like?
Ideas pass through three periods: 
\bei
\item It can't be done. 
\item It probably can be done, but it's not worth doing. 
\item I knew it was a good idea all along!
\eei
From the depths of my memory $\dots$
\begin{remark}
\ccmark It can't be done: at LHC we reconstruct the invariant mass of the Higgs decay products,
``easy" in case of $\PGg\PGg$ or $4$ charged lepton final states. The mass
resolution has a Gaussian core but non-Gaussian tails (\eg, due to
calorimeter segmentation but also pile-up effects \etc). The accuracy
in the mean of the mass peak can then approach that $1.\%$ precision.
Thus it could perhaps compare with the $\PW\,$-mass extraction at LEP,
based on some measured invariant mass distribution. Experimentalists
would let the detector event simulation program do the folding of the
theoretical invariant mass distribution, hoping that the MC catches
most of the Gaussian and non-Gaussian resolution effects with the
remainder being put into the systematic uncertainty. However, this
would affect the width much more than the mass (mean of the distribution).
\end{remark}
\begin{remark}
\ccmark It's not worth doing. 
For the width of the Higgs things are thus much more difficult:
For $\mh < 180\UGeV$ detector resolution dominates, so experimentally it will be very
tough. 
\end{remark}
Let's review what we have learned in the meantime, highlighting new steps for Higgs precision 
physics:
\bei
\item complete off-shell treatment of the Higgs signal 
\item signal-background interference 
\item residual theoretical uncertainty
\eei
\section{The wrath of the ``heavy'' Higgs \label{wrath}}
\begin{center}
\textcursive{You didn't want me to be real, I will contaminate your data,
come and see if ye can swerve me}
\end{center}
Let's see how this develops.
\subsection{Higgs boson Production and decay: the analytic structure\label{Sect_prodec}}
\begin{remark}
\cmark I knew it was a good idea all along!
\end{remark}
Before giving an unbiased description of production and decay of an Higgs boson we underline 
the general structure of any process containing a Higgs boson intermediate state. The 
corresponding amplitude is schematically given by
\bq
A(s) = \frac{f(s)}{s - \cph} + N(s),
\label{presplit}
\eq
where $N(s)$ denotes the part of the amplitude which is non-Higgs-resonant.
Strictly speaking, signal ($\mathrm{S}$) and background ($\mathrm{B}$) should be defined as 
follows:
\bq
A(s) = S(s) + B(s),
\qquad
S(s)= \frac{f(\cph)}{s - \cph}, \quad
B(s)= \frac{f(s) - f(\cph)}{s - \cph} + N(s)
\label{split}
\eq
\begin{definition}
\begin{itshape}
The Higgs complex pole (describing an unstable particle) is conventionally parametrized as
\end{itshape}
\bq
\cph = \mu^2_{\PH} - i\,\mu_{\PH}\,\gamma_{\PH}
\label{CPpar}
\eq
\end{definition}
As a first step we will show how to write $f(s)$ in a way such that pseudo-observables 
make their appearance~\cite{Passarino:2010zza,Passarino:2010qk}. Consider the process 
$i j \to \PH \to \PF$ where $i,j\,\in\,$partons and $\PF$ is a generic final state; the complete 
cross-section will be written as follows:
\bq
\sigma_{i j \to \PH \to \PF}(s) = 
\frac{1}{2\,s}\,\int\,d\Phi_{i j \to \PF}\,
\left[\, \sum_{s,c} \bmid A_{i j \to \PH} \bmid^2\, \right]\,
\frac{1}{\bmid s - \cph\bmid^2}\,
\left[\, \sum_{s,c} \bmid A_{\PH \to \PF} \bmid^2\, \right]
\eq
where $\sum_{s,c}$ is over spin and colors (averaging on the initial state). 
Note that the background (\eg $\Pg\Pg \to 4\,\Pf$) 
has not been included and, strictly speaking and for reasons of gauge invariance, one should 
consider only the residue of the Higgs-resonant amplitude at the complex pole, as described in 
\eqn{split}. 
For gauge invariance the rule of thumb can be formulated by looking at \eqn{presplit}: the only
gauge invariant quantities are the location of the complex pole, its residue and the non-resonant
part of the amplitude ($B(s)$ of \eqn{split}). 
For the moment we will argue that the dominant corrections are the QCD ones 
where we have no problem of gauge parameter dependence. If we decide to keep the Higgs 
boson off-shell also in the resonant part of the amplitude (interference signal/background 
remains unaddressed) then we can write
\bq
\int\,d\Phi_{i j \to \PH}\,\sum_{s,c} \bmid A_{i j \to \PH} \bmid^2 = s\,{\overline A}_{ij}(s).
\eq
For instance, we have
\bq
{\overline A}_{\Pg\Pg}(s) = \frac{\alphas^2}{\pi^2}\,\frac{G_{\ssF}\,s}{288\,\sqrt{2}}\,
\bmid \sum_q\,f\lpar \tau_q\rpar \bmid^2\,\lpar 1 + \delta_{\QCD}\rpar,
\eq
where $\tau_q = 4\,m^2_q/s$, $f(\tau_q)$ is defined in Eq.(3) of \Bref{Spira:1995rr} and where 
$\delta_{\QCD}$ gives the QCD corrections to $\Pg\Pg \to \PH$ up to 
next-to-next-to-leading-order (NNLO) + next-to-leading logarithms (NLL) resummation. 
Furthermore, we define
\bq
\Gamma_{\PH \to \PF}(s) =
\frac{1}{2\,\sqrt{s}}\,\int\,d\Phi_{\PH \to \PF}\,\sum_{s,c} \bmid A_{\PH \to \PF} \bmid^2
\label{offw}
\eq
which gives the partial decay width of a Higgs boson of virtuality $s$ into a final 
state $\PF$.
\bq
\sigma_{i j \to \PH}(s) = \frac{{\overline A}_{ij}(s)}{s}
\eq
which gives the production cross-section of a Higgs boson of virtuality $s$.
We can write the final result in terms of pseudo-observables
\begin{proposition}
The familiar concept of on-shell production$\,\otimes\,$branching ratio can be generalized to
\bq
\sigma_{i j \to \PH \to \PF}(s) = \frac{1}{\pi}\,
\sigma_{i j \to \PH}(s)\,\frac{s^2}{\bmid s - \cph\bmid^2}\,
\frac{\Gamma_{\PH \to \PF}(s)}{\sqrt{s}}
\label{offs}
\eq
It is also convenient to rewrite the result as 
\bq
\sigma_{i j \to \PH \to \PF}(s) = \frac{1}{\pi}\,
\sigma_{i j \to \PH}\,\frac{s^2}{\bmid s - \cph\bmid^2}\,
\frac{\Gamma^{\tot}_{\PH}}{\sqrt{s}}\,\mathrm{BR}\lpar \PH \to \PF\rpar
\label{QFT}
\eq
where we have introduced a sum over all final states,
\bq 
\Gamma^{\tot}_{\PH} = \sum_{\Pf\in\PF}\,\Gamma_{\PH \to \Pf}
\label{Gtot}
\eq
\end{proposition}
Note that we have written the phase-space integral for $i(p_1) + j(p_2) \to \PF$ as
\bq
\int\,d\Phi_{i j \to \PF} =
\int\,d^4k\,\delta^4(k - p_1 - p_2)\,
\int\,\prod_f\,d^4p_f\,\delta^+(p^2_f)\,\delta^4(k - \sum_f p_f)
\eq
where we assume that all initial and final states (\eg $\PGg \PGg$, $4\,\Pf$, \etc) 
are massless. 

Why do we need pseudo-observables? Ideally experimenters (should) extract so-called 
{\em realistic observables} from raw data, \eg 
$\sigma \lpar \Pp\Pp \to \PGg\PGg + \PX\rpar$ and (should) present results in a form that 
can be useful for comparing them with theoretical predictions, i.e. the results should be 
transformed into pseudo-observables; during the deconvolution procedure one should also account 
for the interference background -- signal;
theorists (should) compute pseudo-observables using the best available technology and satisfying 
a list of demands from the self-consistency of the underlying theory.
\begin{definition}
\begin{itshape}
We define an off-shell production cross-section (for all channels) as follows:
\end{itshape}
\bq
\sigma^\prop_{i j \to \ac} =
\frac{1}{\pi}\,\sigma_{i j \to \PH}\,\frac{s^2}{\bmid s - \cph\bmid^2}\,
\frac{\Gamma^{\tot}_{\PH}}{\sqrt{s}}
\label{sigmaPR}
\eq
\end{definition}
When the cross-section $i j \to \PH$ refers to an off-shell Higgs boson the choice of the QCD 
scales should be made according to the virtuality and not to a fixed value. Therefore,
for the PDFs and $\sigma_{i j \to \PH  + \PX}$ one should select $\muFs = \muRs = z\,s/4$ ($z\,s$ 
being the invariant mass of the detectable final state). Indeed, beyond lowest order (LO) one must 
not choose the invariant mass of the incoming partons for the renormalization and factorization 
scales, with the factor $1/2$ motivated by an improved convergence of fixed order expansion, but 
an infrared safe quantity fixed from the detectable final state, see \Bref{Collins:1989gx}. The 
argument is based on minimization of the universal logarithms (DGLAP) and not the process-dependent
ones.
\subsection{More on production cross-section \label{more}}
We give the complete definition of the production cross-section; let us define $\upzeta= z\,s$, 
$\upkappa= v\,s$, and write
\begin{definition}
\begin{itshape}
$\sigma^{\myprod}$ is defined by the following equation:
\end{itshape}
\bq
\sigma^{\myprod} = 
\sum_{i,j}\,\int \hbox{PDF}\,\otimes\,\sigma^{\myprod}_{i j \to \ac} = 
\sum_{i,j}\,\int_{z_0}^1 dz \int_z^1 \frac{dv}{v}\,{\mathcal{L}}_{ij}(v)
\sigma^\prop_{ij \to \ac}(\upzeta, \upkappa, \muR, \muF)
\label{PDFprod_1}
\eq
\begin{itshape}
where $z_0$ is a lower bound on the invariant mass of the $\PH$ decay products,
the luminosity is defined by
\end{itshape}
\bq
{\mathcal{L}}_{ij}(v) = \int_v^1 \frac{dx}{x}\,
f_i\lpar x,\muF\rpar\,f_j\lpar \frac{v}{x},\muF\rpar
\label{PDFprod_2}
\eq
\begin{itshape}
where $f_i$ is a parton distribution function and
\end{itshape}
\bq
\sigma^\prop_{ij \to \ac}(\upzeta, \upkappa, \muR, \muF) = \frac{1}{\pi}\, 
\sigma_{ij \to \PH  + \PX}(\upzeta, \upkappa, \muR, \muF)\,
\frac{ \upzeta\,\upkappa}{\bmid \upzeta - \cph\bmid^2}\,
\frac{\Gamma^{\tot}_{\PH}(\upzeta)}{\sqrt{\upzeta}}
\label{PDFprod_3}
\eq
Therefore, $\sigma_{ij \to \PH  + \PX}(\upzeta, \upkappa, \muR)$ is the cross section for two
partons of invariant mass $\upkappa$ ($z \le v \le 1$) to produce a final state containing 
a $\PH$ of virtuality $\upzeta= z\,s$ plus jets (X); it is made of several 
terms (see \Bref{Spira:1995rr} for a definition of $\Delta\sigma$),
\bq
\sum_{ij}\,\sigma_{ij \to \PH  + \PX}(\upzeta, \upkappa, \muR, \muF) = 
\sigma_{\Pg\Pg \to \PH}\,\delta\lpar 1 - \frac{z}{v}\rpar + \frac{s}{\upkappa}\,\lpar
\Delta\sigma_{\Pg\Pg \to \PH\Pg} + \Delta\sigma_{\PQq\Pg \to \PH\PQq} + 
\Delta\sigma_{\PAQq\PQq \to \PH\Pg} + \mbox{NNLO}\rpar  
\label{PDFprod_4}
\eq
\end{definition}
\begin{remark}
As a technical remark the complete phase-space integral for the process
$\hatp_i + \hatp_j \to p_k + \{f\}$ ($\hatp_i = x_i\,p_i$ \etc) is written as
\bqa
\int\,d\Phi_{ij \to f} &=& 
\int\,d\Phi_{\myprod}\,\int\,d\Phi_{\dec} =
\int d^4 p_k\,\delta^+(p^2_k)\,\prod_{l=1,n} d^4 q_l\,\delta^+(q^2_l)\,
\delta^4\lpar \hatp_i + \hatp_j - p_k - \sum_l q_l\rpar 
\nl
{}&=&
\int d^4k d^4Q\,\delta^+(p^2_k)\,\delta^4 \lpar \hatp_i + \hatp_j - p_k - Q\rpar\, 
\int \prod_{l=1,n} d^4q_l\,\delta^+(q^2_l)\,\delta^4 \lpar Q - \sum_l q_l\rpar
\eqa
where $\int\,d\Phi_{\dec}$ is the phase-space for the process $Q \to \{f\}$ and
\bqa
\int\,d\Phi_{\myprod} &=& s\,\int dz\,\int d^4p_k d^4Q\,\delta^+(p^2_k)\,
\delta\lpar Q^2 - \upzeta\rpar\,\theta(Q_0)\,
\delta^4 \lpar \hatp_i + \hatp_j - p_k - Q\rpar
\nl
{}&=&
s^2\,\int dz dv d{\hat t}\,\int d^4p_k d^4Q\,\delta^+(p^2_k)\,
\delta\lpar Q^2 - \upzeta\rpar\,\theta(Q_0)\,
\delta^4 \lpar \hatp_i + \hatp_j - p_k - Q\rpar
\nl
{}&\times&
\delta\lpar (\hatp_i + \hatp_j)^2 - \upkappa\rpar\,
\delta\lpar (\hatp_i + Q)^2 - {\hat t}\rpar
\eqa
\eqnsc{PDFprod_1}{PDFprod_3} follow after folding with PDFs of argument $x_i$ and $x_j$, after 
using $x_i = x$, $x_j = v/x$ and after integration over ${\hat t}$. At NNLO there is an 
additional parton in the final state and five invariants are need to describe the partonic 
process, plus the $\PH$ virtuality. However, one should remember that at NNLO use is made 
of the effective theory approximation where the Higgs-gluon interaction is described by a 
local operator. 
\end{remark}
\subsection{An idea that is not dangerous is unworthy of being called an idea at all \label{int}}
Let us consider the case of a light Higgs boson; here, the common belief was that the
product of on-shell production cross-section (say in gluon-gluon fusion) and branching
ratios reproduces the correct result to great accuracy. The expectation is based on the well-known
result~\cite{Tkachov:1999qb} ($\Gamma_{\PH} \muchless M_{\PH}$)
\bq
\Delta_{\PH} = \frac{1}{\lpar s - M^2_{\PH}\rpar^2 + \Gamma^2_{\PH}\,M^2_{\PH}} =
\frac{\pi}{M_{\PH}\,\Gamma_{\PH}}\,\delta\lpar s - M^2_{\PH}\rpar +
\mathrm{PV}\,\left[ \frac{1}{\lpar s - M^2_{\PH}\rpar^2}\right] 
\label{CPV}
\eq
where $\mathrm{PV}$ denotes the principal value (understood as a distribution).
Furthermore $s$ is the Higgs virtuality and $M_{\PH}$ and $\Gamma_{\PH}$ should be understood as 
$M_{\PH} = \mu_{\PH}$ and $\Gamma_{\PH} = \gamma_{\PH}$ and not as the corresponding on-shell 
values. In more simple terms, the first term in \eqn{CPV} puts you on-shell and the second one 
gives you the off-shell tail. More details are given in \refA{Aoff}. 
\begin{remark}
\begin{itshape}
$\Delta_{\PH}$ is the Higgs propagator, there is no space for anything else in QFT (\eg 
Breit-Wigner distributions). For a comparison of Breit-Wigner and Complex Pole distributed cross 
sections at $\muh = 125.6\UGeV$ see \refF{fig:HTO_6} and \refF{fig:HTO_6b}.
\end{itshape}
\end{remark}
A more familiar representation of the propagator can be written as follows:
\begin{definition}
\begin{itshape}
with the parametrization of \eqn{CPpar} we perform the well-known transformation
\end{itshape}
\bq
\mBhs = \muhs + \gamma^2_{\PH} 
\qquad
\mu_{\sH}\,{\GOL}_{\PH} = \mBh\,\gh
\label{Bars}
\eq
\begin{itshape}
A remarkable identity follows (defining the Bar-scheme):
\end{itshape}
\bq
\frac{1}{s - \cph} =
\lpar 1 + i\,\frac{{\GOL}_{\PH}}{\mBh}\rpar\,
\lpar s - \mBhs + 
i\,\frac{{\GOL}_{\PH}}{\mBh}\,s \rpar^{-1}
\label{barid}
\eq
showing that the Bar-scheme is equivalent to introducing a running width in the propagator
with parameters that are not the on-shell ones. Special attention goes to the
numerator in \eqn{barid} which is essential in providing the right asymptotic
behavior when $s \to \infty$, as needed for cancellations with contact terms in
$\PV\PV\,$scattering. 
\end{definition}

The natural question is: to which level of accuracy does the ZWA (delta-term only in 
\eqn{CPV}) approximate the full off-shell result given that at $\muh = 125\UGeV$ the on-shell
width is only $4.03\UMeV$?
For definiteness we will consider $ij \to \PH \to \PZ\PZ \to 4\Pl$.
When searching the Higgs boson around $125\UGeV$ one should not care about the region
$M_{\ssZZ} > 2\,M_{\PZ}$ but, due to limited statistics, theory predictions for the
normalization in $\PAQq-\PQq - \Pg\Pg \to  \PZ\PZ$ are used over the entire spectrum in the
$\PZ\PZ$ invariant mass. 

Therefore, the question is not to dispute that off-shell effects are depressed by a factor
$\gh/\muh$ but to move away from the peak and look at the behavior of the invariant 
mass distribution, no matter how small it is compared to the peak; is it really decreasing with 
$M_{\ssZZ}$? Is there a plateau? For how long? How does that affect the total cross-section if
no cut is made?

Let us consider the signal, in the complex-pole scheme:
\bq
\sigma_{\Pg \Pg \to \PZ \PZ}(\mrS) =
\sigma_{\Pg \Pg \to \PH \to \PZ \PZ}(M^2_{\ssZZ}) = \frac{1}{\pi}\,
\sigma_{\Pg \Pg \to \PH}\,\frac{M^4_{\ssZZ}}{\bmid M^2_{\ssZZ} - \cph\bmid^2}\,
\frac{\Gamma_{\PH \to \PZ\PZ}\lpar M_{\ssZ}\rpar}{M_{\ssZZ}}
\eq
\label{signal}
where $\cph$ is the Higgs complex pole, given in \eqn{CPpar}.
Away (but not too far away) from the narrow peak the propagator and the off-shell $\PH$ width 
behave like 
\bq
\Delta_{\PH} \approx \frac{1}{\lpar M^2_{\ssZZ} - \muhs\rpar^2},
\qquad
\frac{\Gamma_{\PH \to \PZ\PZ}\lpar M_{\ssZ}\rpar}{M_{\ssZZ}} \sim G_{\ssF}\,M^2_{\ssZZ}
\eq
above threshold with a sharp increase just below it (it goes from $1.62\,\cdot\,10^{-2}\UGeV$
at $175\UGeV$ to $1.25\,\cdot\,10^{-1}\UGeV$ at $185\UGeV$).

\begin{figure}[h!]
\begin{minipage}{.9\textwidth}
\begin{center}
  \includegraphics[width=1.0\textwidth, bb = 0 0 595 842]{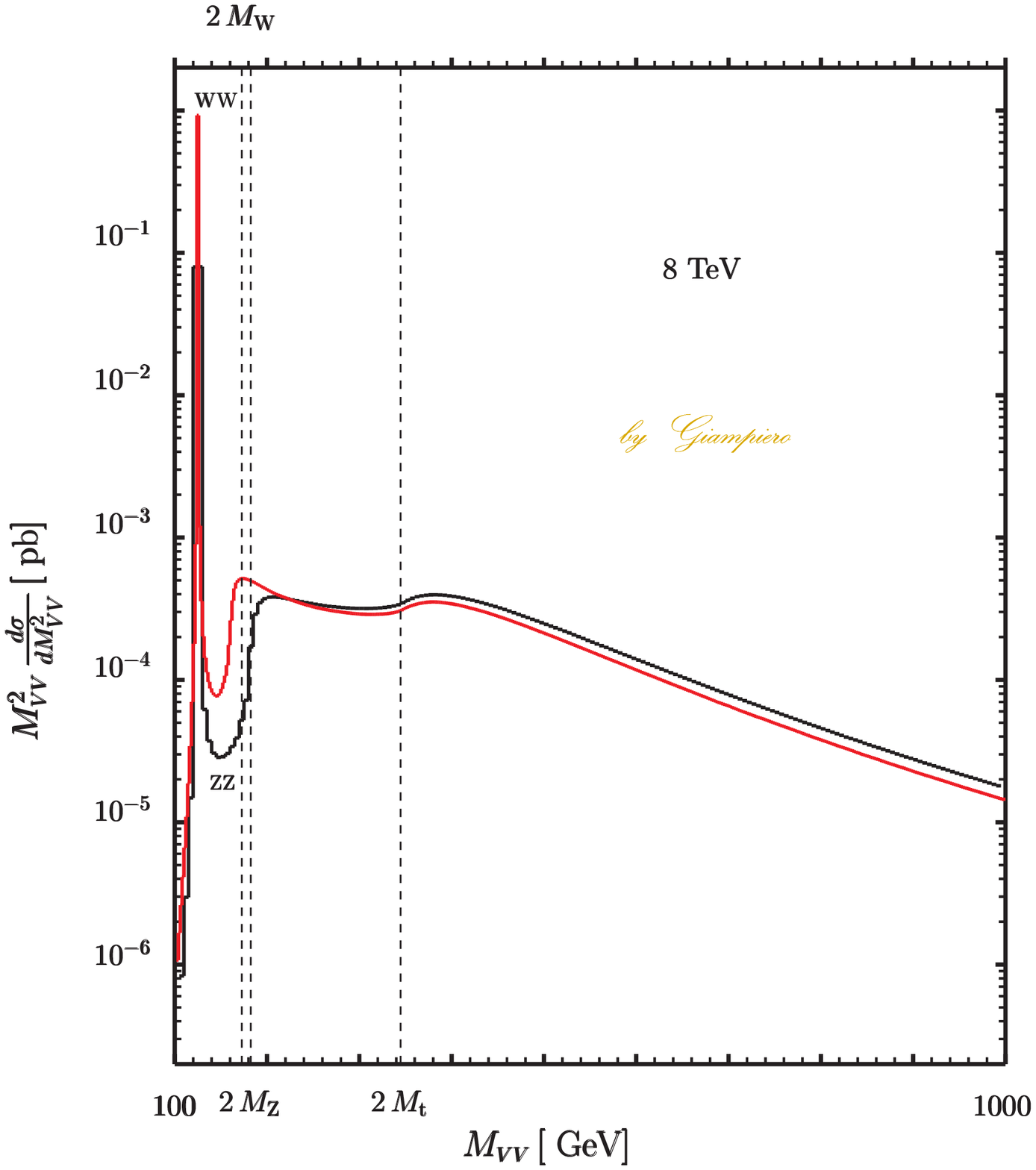}
  \vspace{-3.6cm}
  \caption{
The NNLO $\PV\PV$ invariant mass distribution in $\Pg\Pg \to \PV\PV$ for $\muh = 125\UGeV$.}
\label{fig:HTO_1}
\end{center}
\end{minipage}
\end{figure}

Our result for the $\PV\PV$ ($\PV= \PW/\PZ$) invariant mass distribution is shown in 
\refF{fig:HTO_1}: after the peak the distribution is falling down until the effects of the 
$\PV\PV\,$-thresholds become effective with a visible increase followed by a plateau, by another 
jump at the $\PAQt-\PQt$-threshold. Finally the signal distribution starts again to decrease, 
almost linearly.

What is the net effect on the total cross-section? We show it in \refT{tab:HTO_1} where
the contribution above the $\PZ\PZ\,$-threshold amounts to $7.6\%$. The presence of the
effect does not depend on the propagator function used (Breit-Wigner or complex-pole propagator). 
The size of the effect is related to the distribution function. In \refT{tab:HTO_2} we present 
the invariant mass distribution integrated bin-by-bin.

If we take the ZWA value for the production cross-section at $8\UTeV$ and for 
$\muh= 125\UGeV$ ($19.146\Upb$) and use the branching ratio into $\PZ\PZ$ of 
$2.67\,\cdot\,10^{-2}$ we obtain a ZWA result of $0.5203\Upb$ with a $5\%$ difference 
w.r.t. the off-shell result, fully compatible with the $7.6\%$ effect coming form the high-energy 
side of the resonance.

Always from \refT{tab:HTO_1} we see that the effect is much less evident if we sum over all final
states with a net effect of only $0.8\%$ (the decay is $\PAQb-\PQb$ dominated).

\begin{figure}[h!]
\begin{minipage}{.9\textwidth}
\begin{center}
  \includegraphics[width=1.0\textwidth, bb = 0 0 595 842]{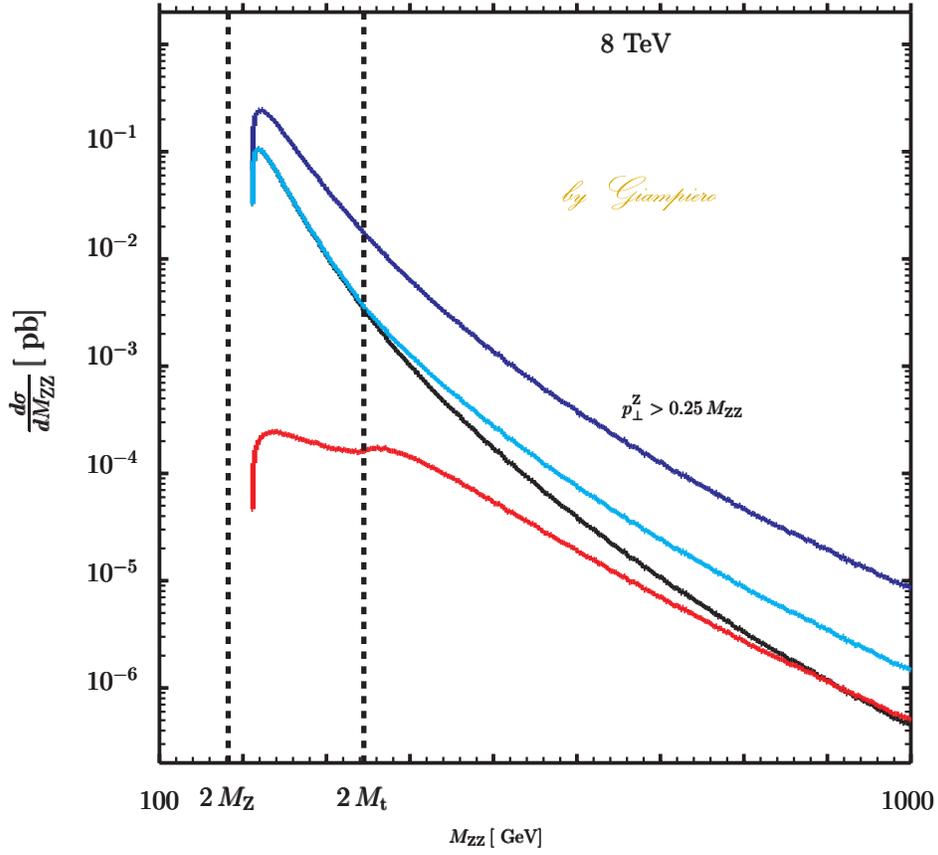}
  \vspace{-3.6cm}
  \caption{
The LO $\PZ\PZ$ invariant mass distribution $\Pg\Pg \to \PZ\PZ$ for $\muh = 125\UGeV$. The 
black line is the total, the red line gives the signal while the cyan line gives signal plus
background; the blue line includes the $q\bar{q} \to \PZ\PZ$ contribution.}
\label{fig:HTO_2}
\end{center}
\end{minipage}
\end{figure}

Of course, the signal per se is not a physical observable and one should always include
background and interference. In \refF{fig:HTO_2} we show the complete LO result. Numbers are 
shown with a cut of $0.25\,M_{\ssZZ}$ on $\pTZ$. The large 
destructive effects of the interference wash out the peculiar structure of the signal
distribution. If one includes the region $M_{\ssZZ} > 2\,M_{\PZ}$ in the analysis then the 
conclusion is: interference effects are relevant also for the low-mass region.

It is worth noting again that the whole effect on the signal has nothing to do with 
$\gh/\muh$ effects; above the $\PZ\PZ\,$-threshold the distribution is higher than
expected (although tiny w.r.t. the narrow peak) and stays approximately constant till the 
$\PAQt-\PQt$-threshold after which we observe an almost linear decrease. This is why the total 
cross-section is affected (in a $\PV\PV$ final state) at the $5\%$ level. 

\begin{table}[hb]
\begin{center}
\caption[]{\label{tab:HTO_1}{
Total cross-section in $\Pg\Pg \to \PH \to \PZ\PZ$ and in $\Pg\Pg \to \PH \to\,$ all;
the part of the cross-section for $M_{\ssZZ} > 2\,M_{\PZ}$ is explicitly shown.}}
\vspace{0.2cm}
\begin{tabular}{cccc}
\hline 
&&&\\
 & Tot[\Upb] & $M_{\ssZZ} > 2\,M_{\PZ}[\Upb]$ & R[\%] \\
&&&\\
$\Pg\Pg \to \PH \to\,$ all  &  $19.146$  & $0.1525$ & $0.8$  \\
$\Pg\Pg \to \PH \to \PZ\PZ$ &  $0.5462$  & $0.0416$ & $7.6$  \\
&&&\\
\hline
\end{tabular}
\end{center}
\end{table}
\begin{table}[t]
\begin{center}
\caption[]{\label{tab:HTO_2}{
Bin-by-bin cross-section in $\Pg\Pg \to \PH \to \PZ\PZ$. First row gives the bin in \UGeV, second 
row gives the cross-section in \Upb.}}
\vspace{0.2cm}
\scalebox{0.9}{
\begin{tabular}{ccccccccc}
\hline 
$\pmb{100-125}$ & $\pmb{125-150}$ & $\pmb{150-175}$ & $\pmb{175-200}$ & $\pmb{200-225}$ & 
$\pmb{225-250}$ & $\pmb{250-275}$ &
$\pmb{275-300}$ \\ 
$\pmb{0.252}$ & $\pmb{0.252}$ & $\pmb{0.195\,\cdot\,10^{-3}}$ & $\pmb{0.177\,\cdot\,10^{-2}}$ &
$\pmb{0.278\,\cdot\,10^{-2}}$ & $\pmb{0.258\,\cdot\,10^{-2}}$ &
$\pmb{0.240\,\cdot\,10^{-2}}$ & $\pmb{0.230\,\cdot\,10^{-2}}$ \\
\hline
\end{tabular}
}
\end{center}
\end{table}
\subsection{When the going gets tough, interference gets going \label{tough}}
The higher-order correction in gluon-gluon fusion have shown a huge $\Kf\,$-factor 
\bq
\Kf = \frac{\sigma_{\myprod}^{\myNNLO}}{\sigma_{\myprod}^{\myLO}},
\qquad
\sigma_{\myprod} = \sigma_{\Pg\Pg \to \PH}.
\eq
\subsubsection{The zero-knowledge scenario \label{zks}}
A potential worry is: should we simply use the full LO calculation or should we try to effectively 
include the large (factor two) $\Kf\,$-factor to have effective NNLO observables? There are 
different opinions since interference effects may be as large or larger than NNLO corrections 
to the signal. Therefore, it is important to quantify both effects. We examine first the scenario
where zero knowledge is assumed on the $\Kf\,$-factor for the background.
So far, two options have been introduced to account for the increase in the signal. Let us 
consider any distribution $\Ds$ (for definiteness we will consider 
$ij \to \PH \to \PZ\PZ \to 4\Pl$),
\ie
\bq
D = \frac{d\sigma}{d M^2_{\ssZZ}} \quad \mbox{or} \quad \frac{d\sigma}{d \pTZ} \quad
\mbox{\etc}
\eq
where $M_{\ssZZ}$ is the invariant mass of the $\PZ\PZ\,$-pair and $\pTZ$ is the
transverse momentum. Two possible options are:
\begin{definition}
\begin{itshape}
The additive option is defined by the following relation
\end{itshape}
\bq
\Ds^{\myNNLO}_{\eff} = \Ds^{\myNNLO}(\mrS) + \Ds^{\myLO}(\mrI) + \Ds^{\myLO}(\mrB)
\label{Aopt}
\eq
\end{definition}
\begin{definition}
\begin{itshape}
The multiplicative~\cite{Campbell:2011cu} ($\mathrm{M}$) or completely multiplicative 
($\overline{\mathrm{M}}$) option is defined by the following relation:
\end{itshape}
\bq
\Ds^{\myNNLO}_{\eff}(\mathrm{M}) = 
\Kf_{\ssD}\,\left[ \Ds^{\myLO}(\mrS) + \Ds^{\myLO}(\mrI) \right] + \Ds^{\myLO}(\mrB),
\quad
\Ds^{\myNNLO}_{\eff}(\overline{\mathrm{M}}) = 
\Kf_{\ssD}\,\Ds^{\myLO},
\quad
\Kf_{\ssD} = \frac{\Ds^{\myNNLO}(\mrS)}{\Ds^{\myLO}(\mrS)}
\label{Mopt}
\eq
where $\Kf_{\ssD}$ is the differential $\Kf\,$-factor for the distribution.
The $\overline{\mathrm{M}}$ option is only relevant for background subtraction and it is
closer to the central value described in \refS{sks}.
\end{definition}
In both cases the NNLO corrections include the NLO electroweak (EW) part, for 
production~\cite{Actis:2008ug} and decay. The EW NLO corrections for $\PH \to \PW\PW/\PZ\PZ \to 
4\Pf$ can reach a $15\%$ in the high part of the tail.
It is worth noting that the differential $\Kf\,$-factor for the $\PZ\PZ\,$-invariant mass 
distribution is a slowly increasing function of $M_{\ssZZ}$ after $M_{\ssZZ} = 2\,M_{\PQt}$, 
going (\eg for $\muh= 125.6\UGeV$) from $1.98$ at $M_{\ssZZ} = 2\,M_{\PQt}$ to $2.11$ 
at $M_{\ssZZ} = 1\UTeV$.

The two options, as well as intermediate ones, suffer from an obvious problem: they are spoiling 
the unitarity cancellation between signal and background for $M_{\ssZZ} \to \infty$. 
Therefore, our partial conclusion is that any option showing an early onset of unitarity 
violation should not be used for too high values of the $\PZ\PZ\,$-invariant mass.   

Therefore, our first prescription in proposing an effective higher-order interference will be
to limit the risk of overestimation of the signal by applying the recipe only in some
restricted interval of the $\PZ\PZ\,$-invariant mass.
This is especially true for high values of $\muh$ where the off-shell effect is large.
Explicit calculations show that the {\em multiplicative} option is better suited for regions with
destructive interference while the {\em additive} option can be used in regions where the effect 
of the interference is positive, \ie we still miss higher orders from the background amplitude
but do not spoil cancellations between signal and background.

Actually, there is an intermediate options that is based on the following observation:
higher-order corrections to the signal are made of several terms,
see \eqn{PDFprod_1}: the partonic cross-section is defined by
\bq
\sum_{ij}\,\sigma_{ij \to \PH  + \PX}(\upzeta, \upkappa, \muR, \muF) = 
\sigma_{\Pg\Pg \to \PH}\,\delta\lpar 1 - \frac{z}{v}\rpar + \frac{s}{\upkappa}\,\lpar
\Delta\sigma_{\Pg\Pg \to \PH\Pg} + \Delta\sigma_{\PQq\Pg \to \PH\PQq} + 
\Delta\sigma_{\PAQq\PQq \to \PH\Pg} + \mbox{NNLO}\rpar  
\label{PDFprod_5}
\eq
From this point of view it seems more convenient to define
\bq
\Kf_{\ssD} = \Kf_{\ssD}^{\Pg\Pg} + \Kf_{\ssD}^{\rest},
\qquad
\Kf_{\ssD}^{\Pg\Pg} = \frac{\Ds^{\myNNLO}\lpar \Pg\Pg \to \PH(\Pg) \to \PZ\PZ(\Pg)\rpar}
{\Ds^{\myLO}\lpar \Pg\Pg \to \PH \to \PZ\PZ\rpar}
\eq
and to introduce a third option
\begin{definition}
\begin{itshape}
The intermediate option is given by the following relation:
\end{itshape}
\bq
\Ds^{\myNNLO}_{\eff} = \Kf_{\ssD}\,\Ds^{\myLO}(\mrS) + \lpar \Kf_{\ssD}^{\Pg\Pg}\rpar^{1/2}\,
\Ds^{\myLO}(\mrI) + \Ds^{\myLO}(\mrB)
\label{Iopt}
\eq
which, in our opinion, better simulates the inclusion of $\Kf\,$-factors at the level of
amplitudes in the zero knowledge scenario (where we are still missing corrections to the continuum 
amplitude).
\end{definition}
\section{There is no free lunch \label{nfl}}
Summary of (Higgs precision physics) milestones without sweeping under the rug the following 
issues:
\bei

\item[\ding{226}] moving forward, beyond ZWA (see \Bref{Kauer:2012hd})\\
\textcursive{don't try fixing something that is already broken in the first place}.

\item[\ding{226}] Unstable particles require complex-pole-scheme (see \Bref{Goria:2011wa}).

\item[\ding{226}] Off-shell + Interferences + uncertainty in $\PV\PV$ production 
(see \Bref{Passarino:2012ri}).

\item[\ding{226}] See also Interference in di-photon channel, 
\Bref{Dixon:2003yb,deFlorian:2013psa}.

\eei
The so-called area method~\cite{Jackson:1975vf} is not so useless, even for a light Higgs boson.
One can use a measurement of the off-shell region to constrain the couplings of the Higgs boson. 
Using a simple cut-and-count method and one scaling parameter (see \eqn{ascal} in \refS{LOMC}), 
existing LHC data should bound the width at the level of $25{-}45$ times the Standard Model 
expectation~\cite{Caola:2013yja,Campbell:2013una}.
\begin{remark}
Chronology and Historical background\\ 
\textcursive{one cannot influence developments beyond telling his 
side of the story. The judgement about originality, importance, impact \etc is of course up 
to others}
\end{remark}
\bei

\item[\ding{227}] Constraining the Higgs boson intrinsic width has been discussed during several
LHC HXSWG meetings (G.~Passarino, LHC HXSWG epistolar exchange,
\eg $10/25/10$ with CMS ``Are you referring to measuring the width according to the area method 
you discuss in your book~\cite{Bardin:1999ak}? That would be interesting to apply if possible'').

\item[\ding{227}] N.~Kauer was the first person who created a plot clearly showing the enhanced 
Higgs tail. It was shown at the $6$th LHC HXSWG meeting\footnote{ 
https://indico.cern.ch/conferenceDisplay.py?ovw=True$\&$confId=182952}.

\item[\ding{227}] N.~Kauer and G.~Passarino (arXiv:1206.4803 [hep-ph]) confirmed the tail and 
provided an explanation for it, starting a detailed phenomenological study, see 
\Bref{Kauer:2012hd} and also \Brefs{Kauer:2013cga,Kauer:2013qba}. 

\item[\ding{227}] Higgs interferometry has been discussed at length in the LHC HXSWG (epistolar 
exchange, \eg on $05/17/13$ ``$\dotsc$ the interference effects could be used to constrain BSM 
Higgs via indirect Higgs width measurement $\dotsc$ there are large visible effects\footnote{See 
R.~Tanaka talk at
http://indico.cern.ch/conferenceTimeTable.py?confId=202554$\#$all.detailed}).
For a comprehensive presentation, see D.~de~Florian talk at ``Higgs Couplings $2013$''\footnote{
https://indico.cern.ch/contributionListDisplay.py?confId=253774}.

\item[\ding{227}] F.~Caola and K.~Melnikov (arXiv:1307.4935 [hep-ph]) introduced the notion of
$\infty\,$-degenerate solutions for the Higgs couplings to SM particles, observed that the 
enhanced tail, discussed and explained in arXiv:1206.4803 [hep-ph], is obviously 
$\gh\,$-independent and that this could be exploited to constrain the Higgs width 
model-independently if there's experimental sensitivity to the off-peak Higgs 
signal~\cite{Caola:2013yja}.
Once you have a model for increasing the width beyond the SM value, \Bref{Caola:2013yja}
turns the observation of \Bref{Kauer:2012hd} into a bound on the Higgs width, within the given
scenario of degeneracy. 

\item[\ding{227}] J.~M.~Campbell, R.~K.~Ellis and C.~Williams (arXiv:1311.3589 [hep-ph]) 
investigated the power of using a matrix element method (MEM) to construct a kinematic discriminant
to sharpen the constraint~\cite{Campbell:2013una} (with foreseeable extensions in 
MEM$@$NLO~\cite{Campbell:2013uha}). MEM-based analysis has been the first to describe a
method for suppressing $\PAQq-\PQq$ background; the importance of his work cannot be 
overestimated. Complementary results from $\PH \to \PW\PW$ in the high transverse mass region are 
shown in \Bref{Campbell:2013wga}.

\item[\ding{227}] This note provides a more detailed description of the theoretical uncertainty 
associated with the camel-shaped and square-root--shaped tails of a light Higgs boson. 

\item[\ding{227}] A similar analysis, performed for the exclusion of a heavy SM Higgs boson, can 
be found in \Bref{Passarino:2012ri} and in \Bref{Campbell:2011cu} with improvements suggested in 
\Bref{Bonvini:2013jha}.

\eei
\subsection{How to use an LO MC? \label{LOMC}}
The MCs used in the analysis are based on LO calculations, some of them include $\Kf\,$-factors 
for the production but all of them have decay and interference implemented at LO.
The adopted solution is external  ``re-weighting'' (\ie re-weighting with results from some 
analytical code), although rescaling exclusive distributions (\eg in the final state leptons) 
with inclusive $\Kf\,$-factors is something that should not be done, it requires (at least) 
a $1{-}1$ correspondence between the two lowest orders. 

\begin{figure}[h!]
\begin{minipage}{.9\textwidth}
\begin{center}
  \includegraphics[width=1.0\textwidth, bb = 0 0 595 842]{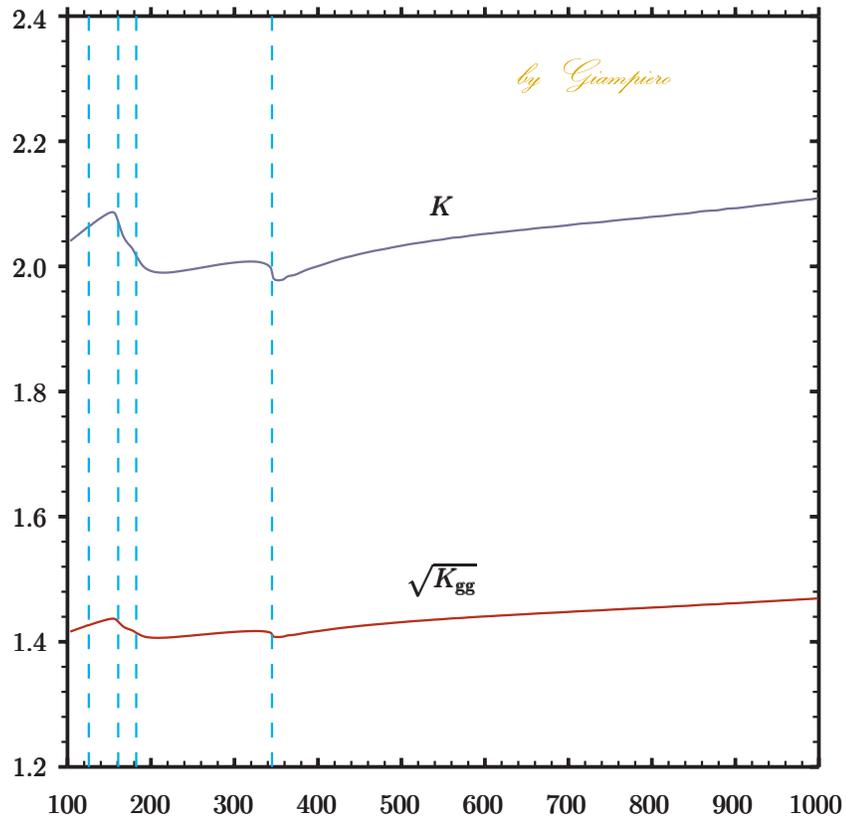}
  \vspace{-3.6cm}
  \caption{
Differential $K\,$-factors in Higgs production for $\muh = 125.6\UGeV$.}
\label{fig:HTO_3}
\end{center}
\end{minipage}
\end{figure}

An example of $\Kf\,$-factors that can be used to include interference in the zero-knowledge 
scenario is given in \refF{fig:HTO_3}. For a more general discussion on re-weighting see 
\Bref{Davatz:2006ut}.

Most of the studies performed so far are for the exclusion of a heavy SM Higgs boson\footnote{cf. 
http://personalpages.to.infn.it/$\tilde{}$giampier/CPHTO.html} and, from that experience, we can
derive that \textcursive{It Takes A Fool To Remain Sane}:
\begin{center}
{\bf{A list of comments and/or problems}}
\end{center}
\bei

\item[\ding{84}] LO decay is not state-of-art, especially for high values of the final state 
invariant mass and the effect of missing higher orders is rapidly increasing with the final state 
invariant mass.

\item[\ding{84}] When the cross-section $i j \to \PH$ refers to an off-shell Higgs boson the 
choice of the QCD scales should be made according to the virtuality and not to a fixed value. 
Indeed, one must choose an infrared safe quantity fixed from the detectable final state, see 
\Bref{Collins:1989gx}. Using the Higgs virtuality or the QCD scales has been advocated in
\Bref{Goria:2011wa}: the numerical impact is relevant, especially for high values of the
invariant mass, the ratio static/dynamic scales being $1.05$. The authors of 
\Bref{Campbell:2013una} seem to agree on our choice~\cite{Goria:2011wa}.

\item[\ding{84}] \Brefs{Caola:2013yja,Campbell:2013una} consider the following scenario (on-shell 
$\infty\,$-degeneracy):
allow for a scaling of the Higgs couplings and of the total Higgs width defined by
\bq
\sigma_{i \to \PH \to f} = \lpar \sigma\cdot\mathrm{BR}\rpar = 
\frac{\sigma^{\myprod}_i\,\Gamma_f}{\gh}
\quad
\sigma_{i \to \PH \to f} \;\varpropto\;\frac{g^2_i g^2_f}{\gh}
\quad 
g_{i,f} = \xi\,g^{\mySM}_{i,f}, \;\; \gh = \xi^4\,\gh^{\mySM}
\label{ascal}
\eq
Looking for $\xi\,$-dependent effects in the highly off-shell region is an approach that raises 
sharp questions on the nature of the underlying extension of the SM; furthermore it does not 
take into account variations in the SM background and the signal strength in $4\Pl$, relative to 
the expectation for the SM Higgs boson, is measured by CMS to be 
$0.91^{+ 0.30}_{-0.24}$~\cite{CMS:xwa} and by ATLAS to be 
$1.43^{+ 0.40}_{-0.35}$~\cite{Aad:2013wqa}.
We adopt the approach of \Bref{LHCHiggsCrossSectionWorkingGroup:2012nn} 
(in particular Eqs.~(1-18)) which is based on the $\upkappa\,$-language, allowing for a consistent 
``Higgs Effective Field Theory'' (HEFT) interpretation, see \Bref{Passarino:2012cb}. 
Negelecting loop-induced vertices, we have
\bq
\frac{\Gamma_{\Pg\Pg}}{\Gamma_{\Pg\Pg}^{\mySM}(\muh)} = 
\frac{\upkappa_{\PQt}^2\cdot\Gamma_{\Pg\Pg}^{\PQt\PQt}(\muh) +
\upkappa_{\PQb}^2\cdot\Gamma_{\Pg\Pg}^{\PQb\PQb}(\muh) +
\upkappa_{\PQt}\upkappa_{\PQb}\cdot
\Gamma_{\Pg\Pg}^{\PQt\PQb}(\muh)}{\Gamma_{\Pg\Pg}^{\PQt\PQt}(\muh) +
\Gamma_{\Pg\Pg}^{\PQb\PQb}(\muh)+\Gamma_{\Pg\Pg}^{\PQt\PQb}(\muh)}
\quad 
\sigma_{i \to \PH \to f} \;=\; 
\frac{\upkappa^2_i \upkappa^2_f}{\upkappa^2_{\PH}}\,\sigma^{\mySM}_{i \to \PH \to f}
\eq
\begin{remark}
\begin{itshape}
The measure of off-shell effects can be interpreted as a constraint on $\gh$ only when we scale 
couplings and total width according to \eqn{ascal} to keep $\sigma_{\peak}$ untouched, although its
value is known with $15{-}20\%$ accuracy.
\begin{proposition}
\begin{scshape}
The generalization of \eqn{ascal} is an $\infty^2\,$-degeneracy, 
$\upkappa_i\,\upkappa_f = \upkappa_{\PH}$. 
\end{scshape}
\end{proposition}
On the whole, we have a constraint in the multidimensional $\upkappa\,$-space, since
$\upkappa^2_{\Pg} = \upkappa^2_{\Pg}(\upkappa_{\PQt},\upkappa_{\PQb})$ and $\upkappa^2_{\PH} = 
\upkappa^2_{\PH}(\upkappa_j,\,\forall\,j)$. Only on the assumption of degeneracy we can prove 
that off-shell effects ``measure'' $\upkappa_{\PH}$; a combination of on-shell effects (measuring 
$\upkappa_i\,\upkappa_f/\upkappa_{\PH}$) and off-shell effects (measuring $\upkappa_i\,\upkappa_f$,
see \eqn{offs}) gives information on $\upkappa_{\PH}$ without prejudices. Denoting by $\mrS$ the 
signal and by $\mrI$ the interference and assuming that $\mrI_{\peak}$ is negligible we have
\bq
\frac{\mathrm{S}_{\off}}{\mathrm{S}_{\peak}}\,\upkappa^2_{\PH} +
\frac{\mathrm{I}_{\off}}{\mathrm{S}_{\peak}}\,\frac{\upkappa_{\PH}}{x_{if}},
\qquad
x_{if} = \frac{\upkappa_i \upkappa_f}{\upkappa_{\PH}}
\eq
for the normalized $\SpI$ off-shell cross section.
\end{itshape}
\end{remark}
The background, \eg $\Pg\Pg \to 4\,\Pl$, is also changed by the inclusion of $d =  6$ operators and
one cannot claim that New Physics is modifying only the signal\footnote{Although one cannot 
disagree with von Neumann ``With four parameters I can fit an elephant, and with five I can make 
him wiggle his trunk''.}. 

\item[\ding{84}] The total systematic error is dominated by theoretical uncertainties, therefore
one should never accept theoretical predictions that cannot provide uncertainty in a systematic way
(\ie providing an algorithm). 

\begin{figure}[h!]
\begin{minipage}{.9\textwidth}
\begin{center}
  \includegraphics[width=1.0\textwidth, bb = 0 0 595 842]{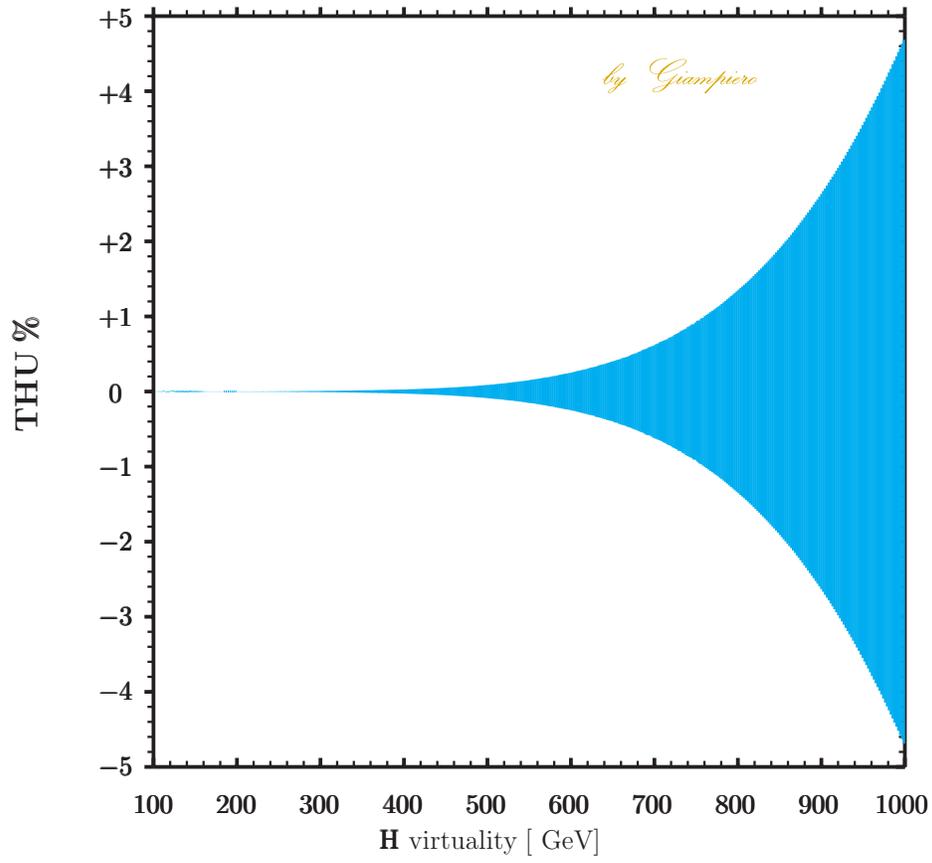}
  \vspace{-3.6cm}
  \caption{
Electroweak theoretical uncertainty for the signal lineshape at $\muh = 125.6\UGeV$}
\label{fig:HTO_5}
\end{center}
\end{minipage}
\end{figure}

\begin{figure}[h!]
\begin{minipage}{.9\textwidth}
\begin{center}
  \includegraphics[width=1.0\textwidth, bb = 0 0 595 842]{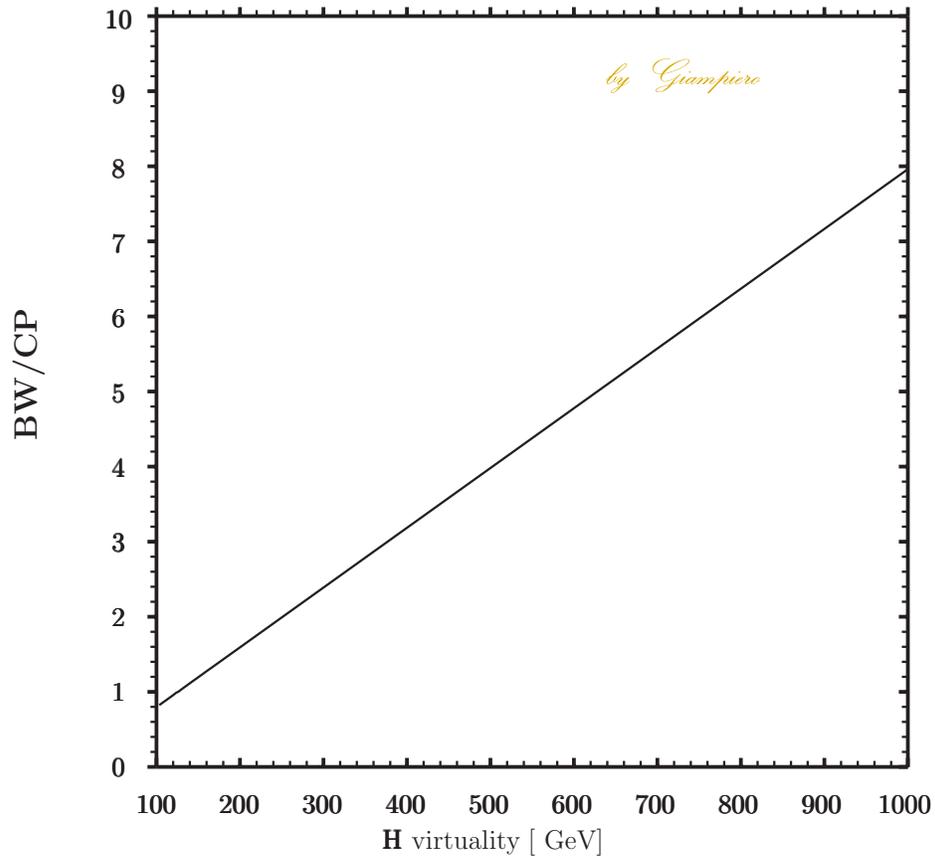}
  \vspace{-3.6cm}
  \caption{
Ratio of Breit-Wigner and Complex Pole distributed cross sections at $\muh = 125.6\UGeV$}
\label{fig:HTO_6}
\end{center}
\end{minipage}
\end{figure}

\begin{figure}[h!]
\begin{minipage}{.9\textwidth}
\begin{center}
  \includegraphics[width=1.0\textwidth, bb = 0 0 595 842]{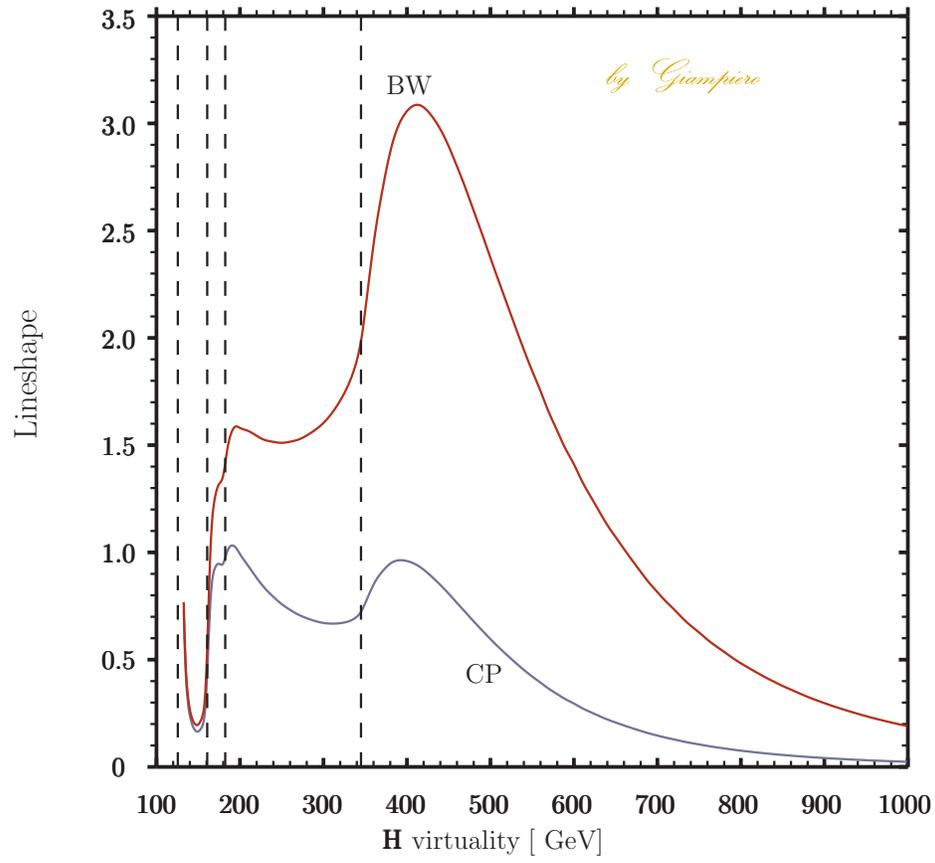}
  \vspace{-3.6cm}
  \caption{
Breit-Wigner and Complex Pole distributed lineshapes at $\muh = 125.6\UGeV$}
\label{fig:HTO_6b}
\end{center}
\end{minipage}
\end{figure}

In \refF{fig:HTO_5} we consider the estimated theoretical uncertainty (THU) on the signal 
lineshape for a mass of $125.6\UGeV$. Note that PDF$\,{+}\,\alphas$ and QCD scales uncertainties 
are not included. As expected for a light Higgs boson, the EW THU is sizable only for large values 
of the off-shell tail, reaching $\pm 4.7\%$ at $1\UTeV$ (the algorithm is explained in 
\Bref{Goria:2011wa}). To summarize the various sources of parametric (PU) and theoretical (THU) 
uncertainties, we have
\begin{center}
{\bf{THU summary}}
\end{center}

\begin{figure}[h!]
\begin{minipage}{.9\textwidth}
\begin{center}
  \includegraphics[width=1.0\textwidth, bb = 0 0 595 842]{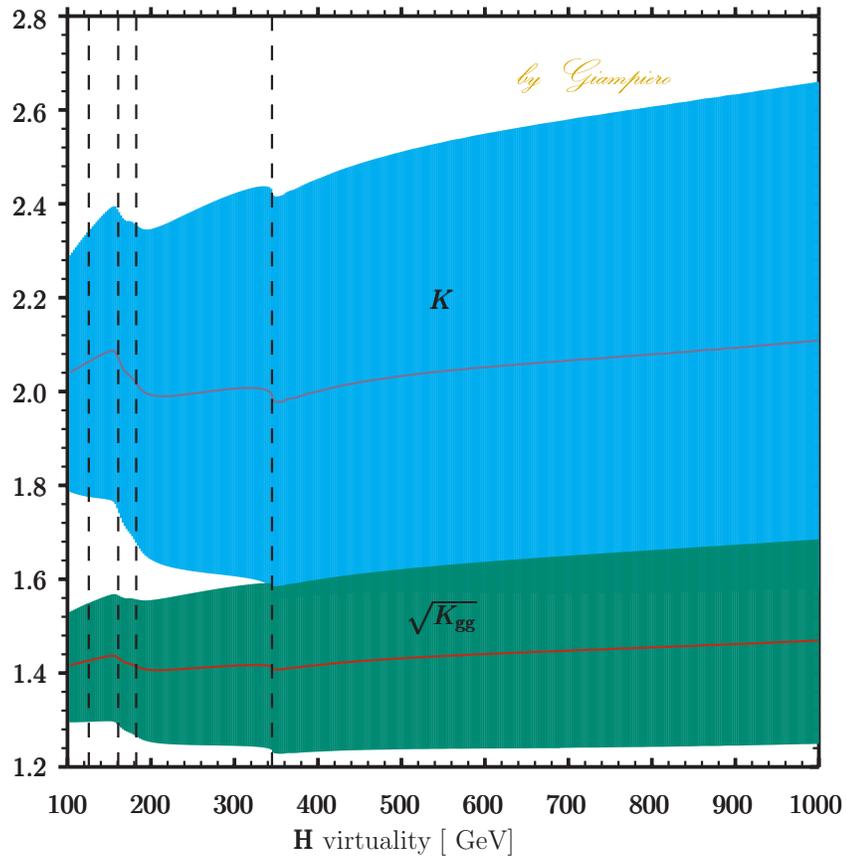}
  \vspace{-3.6cm}
  \caption{
Differential $K\,$-factors in Higgs production for $\muh = 125.6\UGeV$. The central values 
correspond to $\muR = \muF = \mfs/2$, where $\mfs$ is the Higgs virtuality. The bands
give the THU simulated by varying QCD scales $\in\; [ \mfs/4\,,\,\mfs ]$}
\label{fig:HTO_7}
\end{center}
\end{minipage}
\end{figure}

\begin{itemize}

\item[\ding{192}] PDF$\,+\,\alphas$; these have a Gaussian distribution;

\item[\ding{193}] \cmark $\muR, \muF$ (renormalization and factorization QCD scales) variations; 
they are the standard substitute for missing higher order uncertainty (MHOU)~\cite{David:2013gaa}; 
MHOU are better treated in a Bayesian context with a flat prior;

\item[\ding{194}] uncertainty on $\gh$ (\eqn{CPpar}) due to missing higher orders, negligible for 
a light Higgs;

\item[\ding{195}] \cmark uncertainty for $\Gamma_{\PH \to \PF}(\mfs)$ due to missing higher orders 
(mostly EW), especially for high values of the Higgs virtuality $\mfs$ (\ie the invariant mass 
in $\Pp\Pp \to \PH \to \Pf + \PX$); 

\item[\ding{196}] \cmark uncertainty due to missing higher orders (mostly QCD) for the background

\end{itemize}


where \cmark means discussed in this note. When \ding{193} is included one should remember the 
N${}^3$LO effect in gluon-gluon fusion (estimated $+17\%$ in \Bref{Ball:2013bra}) and and 
additional $+7\%$ for an all-order estimate, see \Bref{David:2013gaa}. These numbers refer to 
the fully inclusive $\Kf\,$-factors. The effect of varying QCD scales, 
$\muR = \muF \in [ \mfs/4\,,\,\mfs ]$ is shown in \refF{fig:HTO_7}, for $\Kf$ and 
$\sqrt{\Kf_{\Pg\Pg}}$. 

Once again, it should be stressed that QCD scale variation is only a conventional simulation of the
effect of missing higher orders. Taking \refF{fig:HTO_7} for its face value, we register a 
substantial reduction in the uncertainty when $\Kf\,$-factors are included. For instance, we
find 
$[ {-}12.1\%\,,\,{+}11.0\% ]$ for the NNLO prediction around the peak, 
$[ {-}10.9\%\,,\,{+}9.9\% ]$ around $2\,M_{\PZ}$ and 
$[ {-}9.7\%\,,\,{+}6.6\% ]$ at $1\UTeV$. 
The corresponding LO prediction is 
$[ {-}27.3\%\,,\,{+}12.9\% ]$ around the peak,
$[ {-}29.5\%\,,\,{+}32.1\% ]$ around $2\,M_{\PZ}$ and
$[ {-}38\%\,,\,{+}42\% ]$ at $1\UTeV$.
Note that $\muR$ enters also in the values of $\alphas$. 

\begin{figure}[h!]
\begin{minipage}{.9\textwidth}
\begin{center}
  \includegraphics[width=1.0\textwidth, bb = 0 0 595 842]{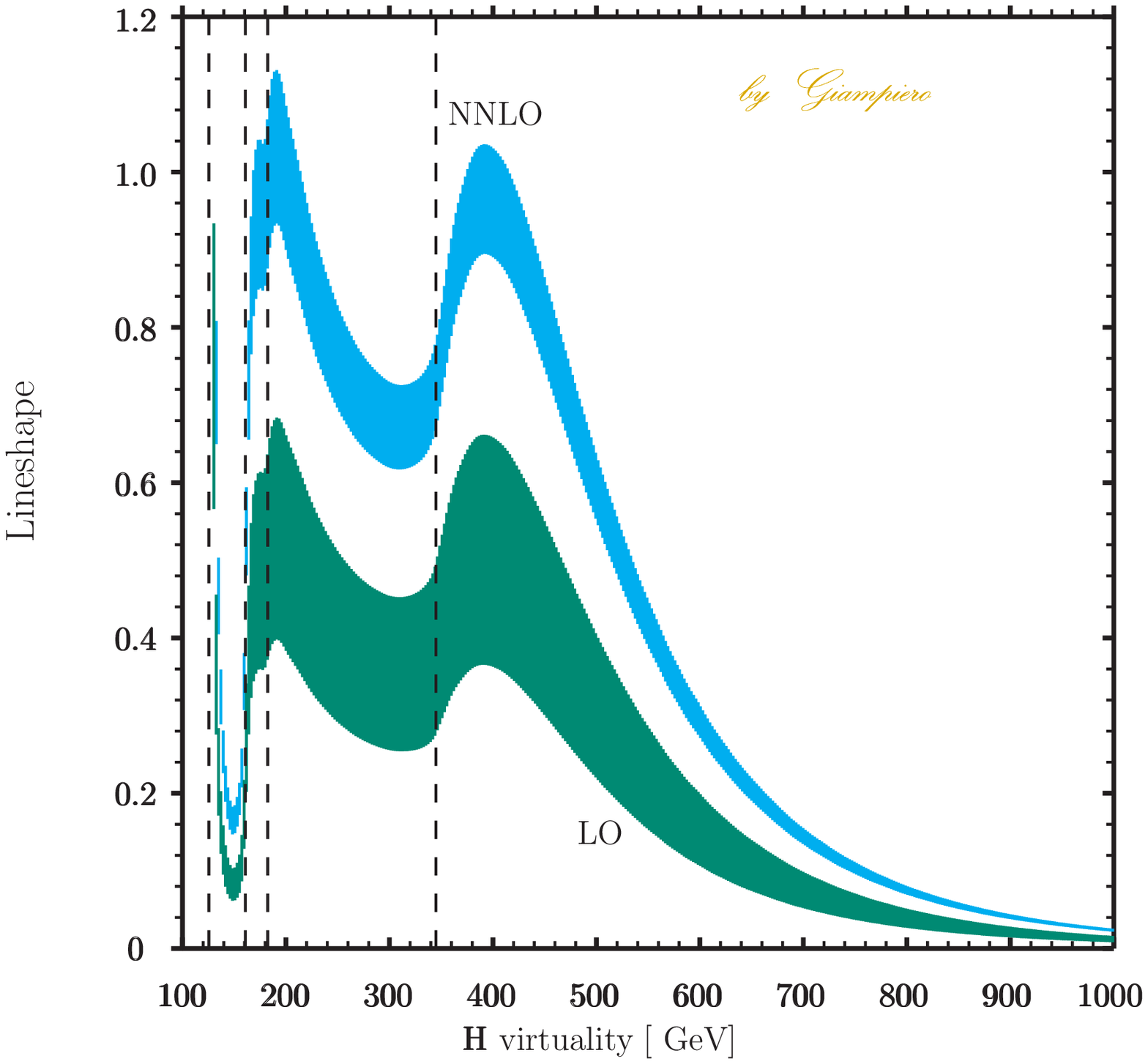}
  \vspace{-3.6cm}
  \caption{
(Camel)Lineshape for $\muh = 125.6\UGeV$. The central values correspond to $\muR = \muF = 
\mfs/2$,  where $\mfs$ is the Higgs virtuality. The bands give the THU simulated by 
varying QCD scales $\in\; [ \mfs/4\,,\,\mfs ]$}
\label{fig:HTO_8}
\end{center}
\end{minipage}
\end{figure}

\begin{figure}[h!]
\begin{minipage}{.9\textwidth}
\begin{center}
  \includegraphics[width=1.0\textwidth, bb = 0 0 595 842]{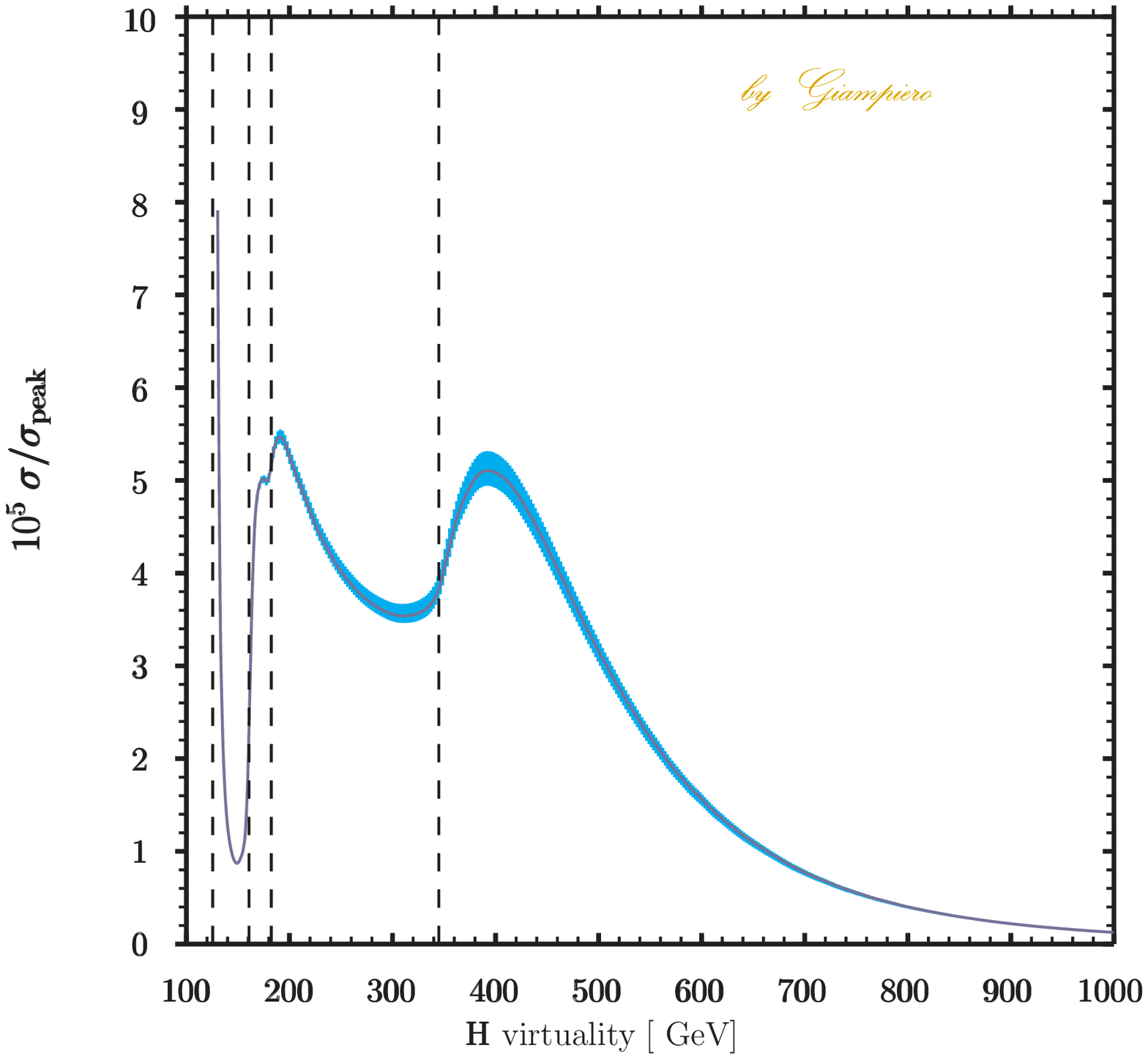}
  \vspace{-3.6cm}
  \caption{
Normalized NNLO lineshape for $\muh = 125.6\UGeV$. The central values correspond to $\muR = \muF = 
\mfs/2$, where $\mfs$ is the Higgs virtuality. The bands give the THU simulated by 
varying QCD scales $\in\; [ \mfs/4\,,\,\mfs ]$}
\label{fig:HTO_9}
\end{center}
\end{minipage}
\end{figure}

Admittedly, showing the effect of QCD scale variations on $\Kf\,$-factors is somewhat misleading 
but we have adopted this choice in view of the fact that, operatively speaking, the experimental 
analysis will generate bins in $M_{4\Pl}$ with a LO MC and multiply the number of events in each 
bin by the corresponding $\Kf\,$-factor. Introducing $\Ds^{\myLO}_{+} = \Ds^{\myLO}(\mfs/4)$ 
and $\Ds^{\myLO}_{-} = \Ds^{\myLO}(\mfs)$, where $\Ds^{\myLO}$ is the LO distribution and
$\Kf_{+} = \Kf(\mfs/4)$ and $\Kf_{-} = \Kf(\mfs)$, where $\Kf = \Ds^{\myNNLO}/\Ds^{\myLO}$ 
is the $\Kf\,$-factor, the correct strategy is $\Kf_{\pm}\,\Ds^{\myLO}_{\pm}$.
When looking at \refF{fig:HTO_7} one should remember that the scale variation that increases
(decreases) the distributions is the one decreasing (increasing) the $\Kf\,$-factor.
The NNLO and LO (camel-shaped) lineshapes, with QCD scale variations, are given in 
\refF{fig:HTO_8}. The THU induced by QCD scale variation can be reduced by considering
the (peak) normalized lineshape, as shown in \refF{fig:HTO_9}.
In other words the constraint on the Higgs intrinsic width should be derived by looking at
the ratio
\bq
\mathrm{R}^{4\,\Pl}_{\off} = 
\frac{\mathrm{N}^{4\,\Pl}_{\off}}{\mathrm{N}^{4\,\Pl}_{\tot}},
\qquad
\mathrm{N}^{4\,\Pl}_{\off} = \mathrm{N}^{4\,\Pl}\lpar M_{4\Pl} > M_0 \rpar
\label{prop}
\eq 
as a function of $\gh/\gh^{\mySM}$, where $\mathrm{N}^{4\Pl}$ is the number of $4\,$-leptons
events. Since the $\Kf\,$-factor has a relatively small range of variation with the virtuality,
the ratio in \eqn{prop} is much less sensitive also to higher order terms. 

An additional comment refers to Eqs.~(42--43) of \Bref{Campbell:2013una}, where $\gh = \gh^{\mySM}$
produces a negative number of events, a typical phenomenon that occurs with large and destructive 
interference effects when only signal + interference is considered. Unless the notion of
negative events is introduced (background-subtracted number of events), the SM case cannot be 
included, as also shown in their Fig.~9, where only the portion $\gh > 4.58(2.08)\,\gh^{\mySM}$ 
should be considered for $M_{4\Pl} > 130(300)\UGeV$, roughly a factor of $10$ smaller than the 
estimated bounds. This clearly demonstrate the importance of controlling THU on the interference,
especially for improved limits on $\gh$.
 
\eei
\subsection{Improving THU for Interference? \label{iTHU}}
One could argue that zero knowledge on the background $\Kf\,$-factor is a too conservative
approach but it should be kept in mind that it's better to be with no one than to be with 
wrong one. Let us consider in details the process $i j \to \PF$; the amplitude can be written as
the sum of a resonant (R) and a non-resonant (NR) part,
\bq
A_{ij \to \PF} = A_{ij \to \PH}\,\frac{1}{s - \cph}\,A_{\PH \to \PF} + A^{\bck}_{ij \to \PF}
\eq
We denote by LO the lowest order in perturbation theory where a process starts contributing and
introduce $\Kf\,$-factors that include higher orders.
\bq
A_{ij \to \PH} = \lpar \Kfac{\mathrm{p}}{ij} \rpar^{1/2}\,A^{\myLO}_{ij \to \PH}, \quad
A_{\PH \to \PF} = \lpar \Kfac{\mathrm{d}}{\PF} \rpar^{1/2}\,A^{\myLO}_{\PH \to \PF}, \quad
A^{\bck}_{ij \to \PF} = \lpar \Kfac{\mathrm{b}}{ij\PF} \rpar^{1/2}\,
A^{\bck,\myLO}_{ij \to \PF}
\label{Kfacts}
\eq
Furthermore, we introduce 
\bq
A^{\res}_{ij \to \PF} = A_{ij \to \PH}\,A_{\PH \to \PF}
\eq
the interference becomes
\bqa
\mathrm{I} &=& 2\,\left[ \Kfac{\mathrm{p}}{ij}\,\Kfac{\mathrm{d}}{\PF}\,
                    \Kfac{\mathrm{b}}{ij\PF} \right]^{1/2}\,
\left\{ \Re\,\frac{A^{\res,\myLO}_{ij \to \PF}}{s - \cph}\;\Re\,A^{\bck,\myLO}_{ij \to \PF} -
        \Im\,\frac{A^{\res,\myLO}_{ij \to \PF}}{s - \cph}\;\Im\,A^{\bck,\myLO}_{ij \to \PF} 
\right\}
\nl
\Re\,\frac{A^{\res,\myLO}_{ij \to \PF}}{s - \cph} &=&
     \frac{s - \muh^2}{\bmid s - \cph\bmid^2}\,\Re\,A^{\res,\myLO}_{ij \to \PF} +
     \frac{\muh\,\gh}{\bmid s - \cph\bmid^2}\,\Im\,A^{\res,\myLO}_{ij \to \PF} 
\nl
\Im\,\frac{A^{\res,\myLO}_{ij \to \PF}}{s - \cph} &=&
     \frac{s - \muh^2}{\bmid s - \cph\bmid^2}\,\Im\,A^{\res,\myLO}_{ij \to \PF} -
     \frac{\muh\,\gh}{\bmid s - \cph\bmid^2}\,\Re\,A^{\res,\myLO}_{ij \to \PF} 
\label{Idet}
\eqa
From \eqn{Idet} we see the main difference in the interference effects of a heavy Higgs boson
w.r.t. the off-shell tail of a light Higgs boson. For the latter case $\gh$ is
completely negligible, whereas it gives sizable effects for the heavy Higgs boson case.
\subsubsection{The soft-knowledge scenario \label{sks}}
Neglecting PDF + $\alphas$ uncertainties and those coming from missing higher orders, the major 
source of THU is due to the missing NLO interference. In \Bref{Bonvini:2013jha} the effect of QCD 
corrections to the signal-background interference at the LHC has been studied for a heavy Higgs 
boson. A soft-collinear approximation to the NLO and NNLO corrections is constructed for the 
background process, which is exactly known only at LO. 
Its accuracy is estimated by constructing and comparing the same approximation to the exact result 
for the signal process, which is known up to NNLO, and the conclusion is that one can describe 
the signal-background interference to better than ten percent accuracy for large values of
the Higgs virtuality. 
It is also shown that, in practice, a fairly good approximation to higher-order QCD corrections 
to the interference may be obtained by rescaling the known LO result by a $\Kf\,$-factor 
computed using the signal process. 

The goodness of the approximation, when applied to the signal, remains fairly good down to 
$180\UGeV$ and rapidly deteriorates only below the $2\,M_{\PZ}\,$-threshold; note that both
$M_{4\Pl} > 130\UGeV$ and $M_{4\Pl} > 300\UGeV$ have been considered in the study of
\Bref{Campbell:2013una}. The exact result for the background is missing but the eikonal nature of 
the approximation should make it equally good, for signal as well as for 
background\footnote{S.~Forte, private communication}.

This line of thought looks very promising, with a reduction of the corresponding THU 
(zero-knowledge scenario), although its extension from the heavy Higgs scenario to the light Higgs 
off-shell scenario has not been completely worked out. In a nutshell, one can write
\bq
\sigma = \sigma^{\myLO} + \sigma^{\myLO}\,\frac{\alphas}{2\,\pi}\,\left[
\mbox{universal} \;+\, \mbox{process dependent} \;+\; \mbox{reg} \right]
\eq
where ``universal'' (the $+$ distribution) gives the bulk of the result while ``process dependent''
(the $\delta$ function) is known up to two loops for the signal but not for the background and
``reg'' is the regular part. A possible strategy would be to use for background the same 
``process dependent'' coefficients and allow for their variation within some ad hoc factor. 
Assuming
\bq
\Kfac{\mathrm{b},\mysoft}{ij\PF} = \Kfac{\mathrm{p}}{ij} \pm \Delta\Kfac{\pm}{ij}
\eq
we could write
\bq
\mathrm{I} = 2\,\Kfac{\mathrm{p}}{ij}\,
\lpar \Kfac{\mathrm{d}}{\PF} \rpar^{1/2}\,
\left[ 1 \pm \frac{\Delta\Kfac{\pm}{ij}}{\Kfac{\mathrm{p}}{ij}} \right]^{1/2}\,
\Re\,\frac{A^{\res,\myLO}_{ij \to \PF}}{s-\cph}\,\lpar A^{\bck\,\myLO}_{ij \to \PF}\rpar^{\ast}
= 2\,\Kfac{\mathrm{p}}{ij}\,
\lpar \Kfac{\mathrm{d}}{\PF} \rpar^{1/2}\,
\left[ 1 \pm \frac{\Delta\Kfac{\pm}{ij}}{\Kfac{\mathrm{p}}{ij}} \right]^{1/2}\,\mathrm{I}^{\myLO}
\eq
In this scenario the subtraction of the background cannot be performed at LO.
It is worth noting that simultaneous inclusion of higher order corrections for Higgs production 
(NNLO) and Higgs decay (NLO) is a three-loop effect that is not balanced even with the 
introduction of the eikonal QCD $\Kf\,$factor for the background; three loop mixed EW-QCD 
corrections are still missing, even at some approximate level. Note that $\Kfac{\mathrm{d}}{4\Pl}$ 
can be obtained by running {\sc Prophecy4f}~\cite{Bredenstein:2007ec} in LO/NLO modes.
\subsection{Background-subtracted lineshape \label{BSL}}
\begin{figure}[h!]
\begin{minipage}{.9\textwidth}
\begin{center}
  \includegraphics[width=1.0\textwidth, bb = 0 0 595 842]{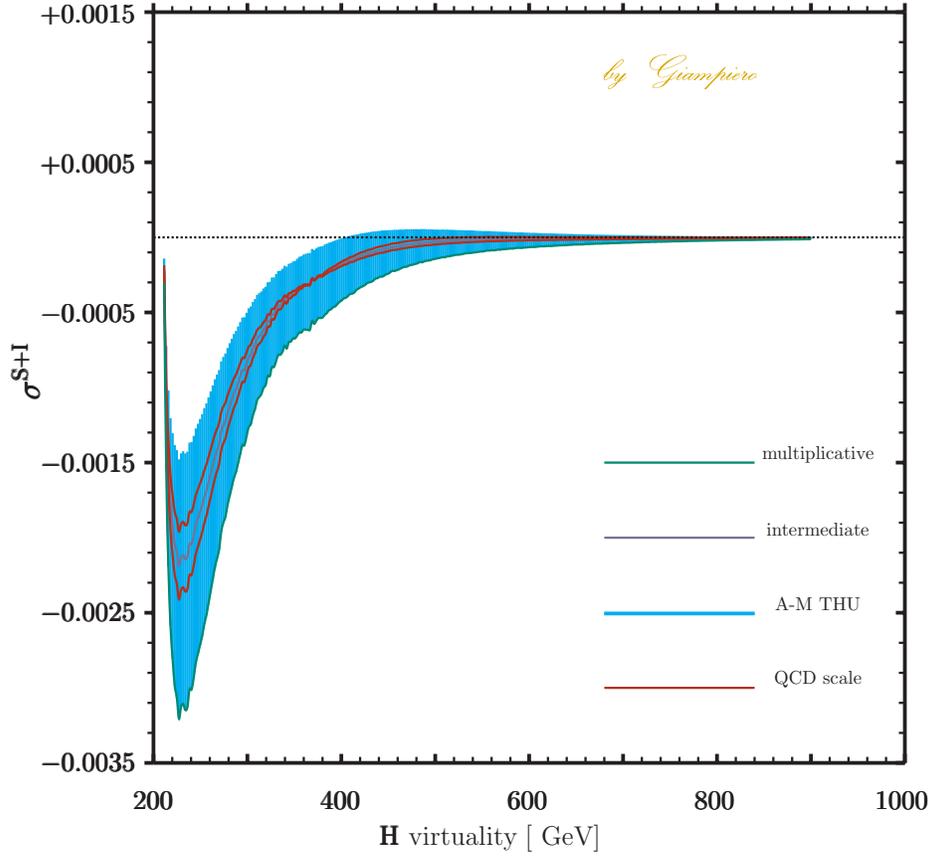}
  \vspace{-3.6cm}
  \caption{
$\sigma^{\SpI}$ for $4\Pe$ final state. The blue curve gives the intermediate
option and the cyan band the associated THU between additive and multiplicative options. 
Multiplicative option is the green curve. Red curves give the THU due to QCD scale variation for 
the intermediate option (QCD scales $\in\; [ \mfs/4\,,\,\mfs ]$, where $\mfs = 
M_{4\Pe}$ is the Higgs virtuality). A cut $\pTZ > 0.25\,M_{4\Pe}$ has been applied.
If one adopts the soft-knowledge recipe, the result is given by the green curve; provisionally, 
one could assume a $\pm 10\%$ uncertainty, extrapolating the estimate made for the high-mass 
study in \Bref{Bonvini:2013jha}}
\label{fig:HTO_10}
\end{center}
\end{minipage}
\end{figure}

In \refF{fig:HTO_10} we present our results for $\sigma^{\SpI}$ for the
$\PZ\PZ \to 4\,\Pe$ final state. The pseudo-observable $\sigma^{\SpI}$ that includes only signal 
and interference (not constrained to be positive) is now a standard in the experimental analysis.

The blue curve in \refF{fig:HTO_10} gives the intermediate option for including the
interference and the cyan band the associated THU between additive and multiplicative options. 
Multiplicative option is the green curve. Red curves give the THU due to QCD scale variation for 
the intermediate option (QCD scales $\in\; [ \mfs/4\,,\,\mfs ]$, where $\mfs = 
M_{4\Pe}$ is the Higgs virtuality). A cut $\pTZ > 0.25\,M_{4\Pe}$ has been applied.
The figure shows how a $\mathrm{S}$ (camel-shaped) distributions transforms into a 
$\SpI$ (square-root--shaped) distribution. 

\begin{remark}
\begin{itshape}
Of course, one could adopt the soft-knowledge recipe, in which case the result is given by
the green curve in \refF{fig:HTO_10}; provisionally, one could assume a $\pm 10\%$ uncertainty,
extrapolating the estimate made for the high-mass study in \Bref{Bonvini:2013jha}.
Background subtraction should be performed accordingly ($\Kfac{\mathrm{b}}{ij\PF}$ of 
\eqn{Kfacts}).
\end{itshape}
\end{remark}

It is worth introducing few auxiliary quantities~\cite{Passarino:2012ri}: the minimum and
the half-minima of $\sigma^{\SpI}$: given
\bq
\Ds\lpar M_{4\,\Pl}\rpar = \frac{d}{d M^2_{4\Pl}}\,\sigma^{\SpI}
\eq
we define
\bq
\Ds_1 = \Ds\lpar M_1\rpar = \min\,\Ds\lpar M_{4\,\Pl}\rpar,
\quad
\Ds^{\pm}_{1/2} = \Ds\lpar M^{\pm}_{1/2}\rpar = \frac{1}{2}\,\Ds\lpar M_1 \rpar
\label{minima}
\eq
As observed in \Bref{Passarino:2012ri}, THU is tiny on $M_1$ and moderately larger for
$M^{\pm}_{1/2}$. 
\begin{remark}
\begin{itshape}
Alternatively, and taking into account the indication of \Bref{Bonvini:2013jha} we could proceed
as follows\footnote{I gratefully acknowledge the suggestion by S.~Bolognesi.}: we can try to turn 
our three {\em measures} of the lineshape into a continuous estimate in each bin; there is a 
technique, called ``vertical morphing''~\cite{Conway:2011in}, that introduces a ``morphing'' 
parameter $f$ which is nominally zero and has some uncertainty. If we define
\end{itshape}
\bq
\Ds^0 = \frac{d \sigma^{\SpI}}{d M^2_{4\Pl}}, \quad \mbox{option}\;\mathrm{I},
\qquad
\Ds^+ = \max_{\ssA,\ssM}\,\Ds, 
\quad
\Ds^- = \min_{\ssA,\ssM}\,\Ds 
\eq
\begin{itshape}
the simplest ``vertical morphing'' replaces
\end{itshape}
\bq
\Ds^0 \to \Ds^0 + \frac{f}{2}\,\lpar \Ds^+ - \Ds^-\rpar
\eq
\begin{itshape}
Of course, the whole idea depends on the choice of the distribution for $f$, usually 
Gaussian which is not necessarily our case; instead, one would prefer to maintain, as much as
possible, the indication from the soft-knowledge scenario (in a Bayesian sense). Therefore, 
we define two curves
\end{itshape}
\bq
\Ds_{-}\lpar \uplambda\,,\,M_{4\Pl}\rpar= \uplambda\,\Ds_{\ssM}\lpar M_{4\Pl}\rpar + 
\lpar 1 - \uplambda\rpar\,\Ds_{\ssI}\lpar M_{4\Pl}\rpar
\qquad
\Ds_{+}\lpar \uplambda\,,\,M_{4\Pl}\rpar= \uplambda\,\Ds_{\ssI}\lpar M_{4\Pl}\rpar + 
\lpar 1 - \uplambda\rpar\,\Ds_{\ssA}\lpar M_{4\Pl}\rpar
\label{moprph}
\eq
\begin{itshape}
We assume that the parameter $\uplambda$, with $0 \le \uplambda \le 1$, has a flat distribution.
We will have $\Ds_{-} < \Ds_{\ssI} < \Ds_{+}$ and a value for $\uplambda$ close to one (\eg $0.9$) 
gives less weight to the additive option, highly disfavored by the eikonal approximation. The 
corresponding THU band will be labelled by $\mathrm{VM}(\uplambda)$.
\end{itshape}
\end{remark}
Consider $\Ds_1$ of \eqn{minima}: we have $M_1 = 233.9\UGeV$ and the THU band corresponding 
to the full variation between A-option and M-option is $0.00171\Ufb$, equivalent to a $\pm 39.9\%$.
If we select $\uplambda = 0.9$ in \eqn{moprph} the difference $\Ds_- - \Ds_+$ 
reduces the uncertainty to $0.00098\Ufb$, equivalent to $\pm 22.8\%$.  
The destructive effect of the interference shows how challenging will be to put more stringent 
bounds on $\gh$ when $\gh \to \gh^{\mySM}$. The off-shell effects are an ideal place where to
look for ``large'' deviations from the SM (from $\gh^{\mySM}$) where, however, large scaling
of the Higgs couplings raise severe questions on the structure of underlying BSM theory.

\begin{figure}[h!]
\begin{minipage}{.9\textwidth}
\begin{center}
  \includegraphics[width=1.0\textwidth, bb = 0 0 595 842]{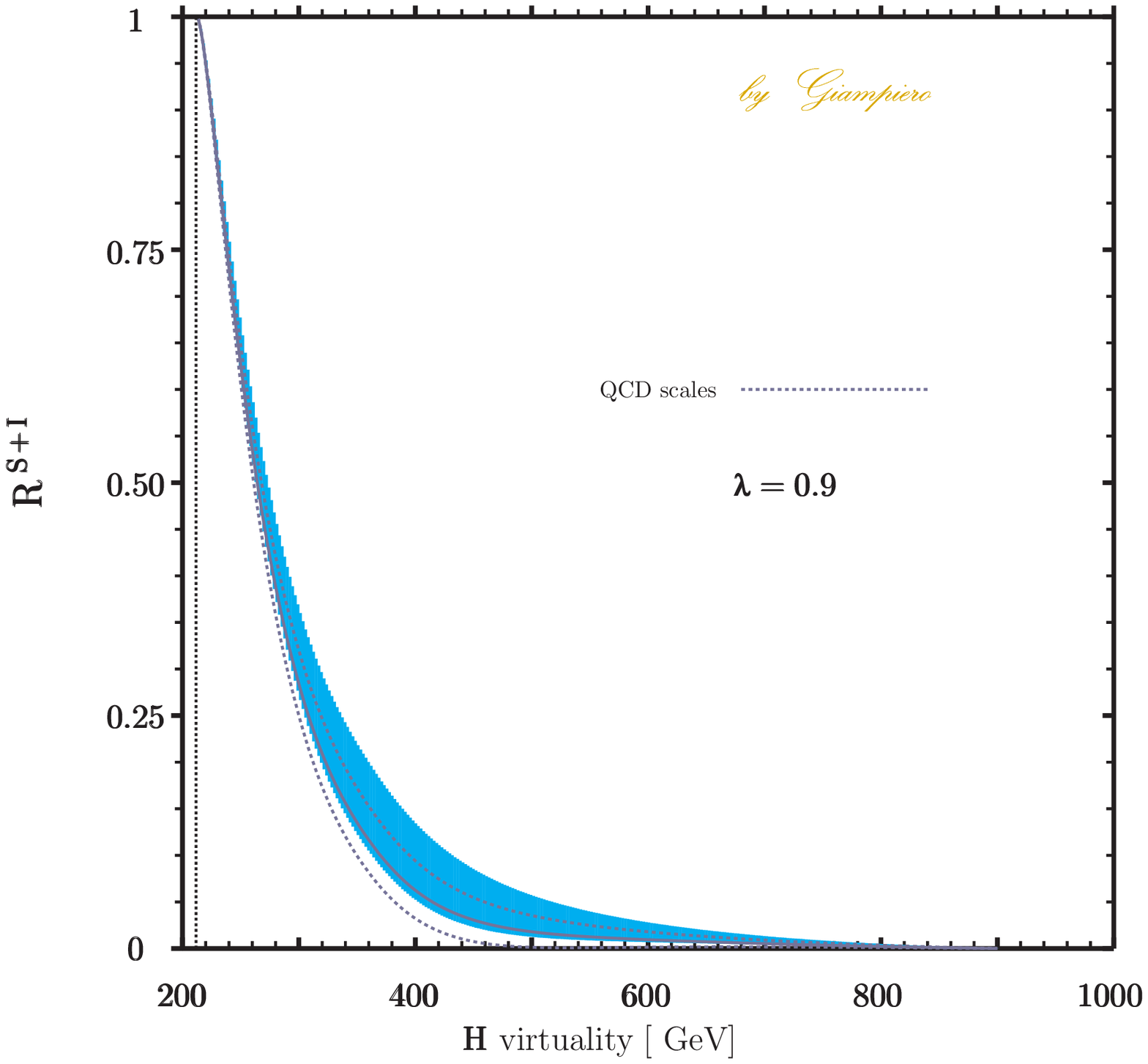}
  \vspace{-3.6cm}
  \caption{
The ratio $\mathrm{R}^{\SpI}(i) = \sigma^{\SpI}(i)/\sigma^{\SpI}(1)$, \eqn{SpI}, where 
$\sigma^{\SpI}(i)$ is obtained by integrating $d\sigma^{\SpI}/d M^2_{4\Pl}$ over bins of 
$2.25\UGeV$ for $M_{4\Pl} > 212\UGeV$. The parameter $\uplambda$ is defined in \eqn{moprph}. 
Dashed lines give the QCD scale variation (QCD scales $\in\; [ \mfs/4\,,\,\mfs ]$, where 
$\mfs = M_{4\Pe}$ is the Higgs virtuality). A cut $\pTZ > 0.25\,M_{4\Pe}$ has been applied}
\label{fig:HTO_11}
\end{center}
\end{minipage}
\end{figure}

\begin{definition}
\begin{itshape}
There is an additional variable that we should consider: 
\end{itshape}
\bq
\mathrm{R}^{\SpI}\lpar M_1,M_2\rpar =
\frac{\sigma^{\SpI}\lpar M_{4\Pl} > M_1 \rpar}{\sigma^{\SpI}\lpar M_{4\Pl} > M_2 \rpar}
\label{SpI}
\eq
\end{definition}
For instance, integrate $d\sigma^{\SpI}/d M^2_{4\Pl}$ over bins of $2.25\UGeV$ for 
$M_{4\Pl} > 212\UGeV$ and obtain $\sigma^{\SpI}(i)$. Next, consider the ratio
$\mathrm{R}^{\SpI}(i) = \sigma^{\SpI}(i)/\sigma^{\SpI}(1)$ which is shown in \refF{fig:HTO_11} 
where the THU band is given by $\mathrm{VM}(0.9)$. To give an example the THU corresponding to 
the bin of $300\UGeV$ is $14.9\%$. THU associated with QCD scale variations is given by the two 
dashed lines.

\section{Conclusions}
The successful search for the on-shell Higgs-like boson has put little emphasis on the potential 
of the off-shell events; the attitude was ``the issue of the Higgs off-shellness is very 
interesting but it is not relevant for low Higgs masses'' and ``for SM Higgs below $200\UGeV$, 
the natural width (mostly for MSSM as well) is much below the experimental resolution. We have 
therefore never cared about it for light Higgs. Just produce on-shell Higgs and let them decay 
in MC''; luckily the panorama is changing.
It is clear that one can't do much without a MC, therefore the analysis should be based on some 
LO MC, or some other. However, more inclusive NLO (or even NNLO) calculations show that  
the LO predictions can be far away, which means that re-weighting can be a better approximation, 
as long as it is accompanied by an algorithmic formulation of the associated theoretical
uncertainty. The latter is (almost) dominating the total systematic error and precision
Higgs physics requires control of both systematics, not only the experimental one.
Very often THU is nothing more than educated guesswork but a workable falsehood is 
more useful than a complex incomprehensible truth. In other words, closeness to the whole truth 
is in part a matter of degree of informativeness of a proposition.
\Acknowledgments
I gratefully acknowledge several important discussions with my inner circle.

\clearpage
\appendix
\section{Appendix: Analytic separation of off-shell effects \label{Aoff}}
The effect of non-SM Higgs couplings on $\sigma^{\SpI}$ can be computed under the assumption
$\gh \muchless \muh$. Consider the following integral:
\bq
F_{ij} =\int_{z_0}^1 dz \int_z^1 \frac{dv}{v}\,{\mathcal{L}}_{ij}(v)\,\bmid f_{ij}(s,z,v)\bmid^2
\label{osplit}
\eq
where the amplitude $f$ is
\bq
f_{ij}(s,z,v) = \frac{A_{ij}(z,v)}{z\,s - \cph} + B_{ij}(z,v)
\eq
and where $ij$ denotes $\Pg\Pg$ or $\PAQq\Pq$. For the process $ij \to \PF$ we have
$A_{ij} \varpropto g_{ij\PH}\,g_{\PH\PF}$ and $A_{iJ}$ is related to
$\sigma_{i j \to \PH}$, $\Gamma_{\PH \to \PF}$ by \eqn{offs}. Simple expressions can be derived 
if we neglect the dependence of $A_{ij}, B_{ij}$ on the kinematic variables (but both contain 
thresholds). Using instead the results of \Bref{Tkachov:1999qb} ($\gh \muchless \muh$) we obtain
\bq
\frac{1}{\bmid z\,s - \cph\bmid^2} =
\frac{\pi}{\muh\,\gh}\,\delta\lpar z\,s - \muhs\rpar +
\mathrm{PV}\,\left[ \frac{1}{\lpar z\,s - \muhs\rpar^2}\right],
\quad
\mathrm{PV}\,\lpar\frac{1}{z^n}\rpar = \frac{(-1)^{n-1}}{(n-1)\,!}\,\frac{d^n}{d z^n}\,
\ln\lpar \mid z\mid\rpar
\eq
we introduce $\omuhs= \muhs/s$, $z_{\PH} = z + \omuhs$ and
\bq
\mathcal{F}^{\ssS}_{ij}(z,v)= \bmid A_{ij}\lpar z,v \rpar\bmid^2, 
\quad
\mathcal{F}^{\ssB}_{ij}(z,v)= \bmid \lpar z\,s - \cph\rpar\,B_{ij}(z,v)\bmid^2,
\quad
\mathcal{F}^{\ssI}_{ij}(z,v)= \lpar z\,s - \cph^{\ast}\rpar\,A_{ij}\lpar z,v \rpar\,
                               B^{\ast}_{ij}\lpar z,v \rpar 
\eq
obtaining the following result for the off-shell part of the integral in \eqn{osplit} 
($z_0 > \omuhs$)
\bq
F_{\off} = -\frac{1}{s^2}\,\int_{z_0}^1\,\frac{d v}{v}\,{\mathcal{L}}_{ij}(v)\,
\int_{z_0-\omuhs}^{v-\omuhs}\,dz
\left[
\mathcal{F}^{\ssS}_{ij}\lpar z_{\PH},v\rpar +
\mathcal{F}^{\ssB}_{ij}\lpar z_{\PH},v\rpar +
2\,\Re\,\mathcal{F}^{\ssI}_{ij}\lpar z_{\PH},v\rpar \right]\,\frac{d^2}{d z^2}\,\ln z
\eq
Since~\cite{Nekrasov:2003er}
\bq
\int_a^b\,dz g(z)\frac{d^2}{d z^2}\,\ln z =
\left[ \frac{g(z)}{z} - g'(z)\,\ln z \right]\,\bmid_a^b + \int_a^b\,dz\,g''(z)\,\ln z
\eq
we derive that the exact behavior of $F_{\off}$ is controlled by the amplitude and by its first two
derivatives. The form factors $\mathcal{F}^l$ admit a formal expansion in $\alphas$ given by 
\bq
\mathcal{F}^l_{ij}(z,v) = \mathcal{F}^{l,0}_i(z)\,\delta\lpar 1 - \frac{z}{v}\rpar +
\sum_{n=1}^{\infty}\,\lpar\frac{\alphas(\muR)}{\pi}\rpar^n\,
\mathcal{F}^{l,n}_{ij}(z,v)
\eq
where we have considered QCD corrections but not the EW.

\clearpage

\clearpage

\clearpage

\clearpage

\clearpage

\clearpage

\clearpage

\clearpage

\clearpage

\clearpage

\clearpage

\clearpage
\bibliographystyle{atlasnote}
\bibliography{HCAT}{}

\end{document}